\documentclass[12pt, a4paper]{article}
\usepackage[height=25cm,width=16cm]{geometry}
\usepackage{feynmp}
\usepackage{amsmath}
\usepackage{ascmac}
\usepackage{dcolumn}
\usepackage{bm,here}
\usepackage{subfig}
\usepackage{comment}
\usepackage{ifpdf}
\usepackage{slashed}
\usepackage{colortbl}
\usepackage{color}
\usepackage{ulem}
\usepackage{comment}
\usepackage{braket}
\usepackage[mathscr]{eucal}
\usepackage[sort&compress, numbers, merge]{natbib}

\usepackage{cancel}

\ifpdf
\usepackage{graphicx}
\usepackage[bookmarksopen,colorlinks=true,linkcolor=bblue,citecolor=ppink,urlcolor=ppink]{hyperref}
\else     % For (p)LaTeX + dvipdfmx
\usepackage[dvipdfmx]{graphicx}  
\usepackage[dvipdfmx,bookmarksopen,colorlinks=true,linkcolor=bblue,citecolor=ppink,urlcolor=ppink]{hyperref}
\fi

\usepackage{multicol}
\definecolor{red}{rgb}{1,0,0}
\definecolor{ppink}{rgb}{0.921545,0.440586,0.687243}
\definecolor{bblue}{rgb}{0.400000,0.400000,1.000000}
\usepackage[charter]{mathdesign}
\usepackage{soul}
\usepackage{wrapfig}

\newcommand{\ditto}{\raisebox{-0.3ex}{\textquotedbl}}

\begin{document}

%%%%%%%%%%%%%%%%%%%%%%%%%%%%%
%%%%%%%%%%% Title %%%%%%%%%%%
%%%%%%%%%%%%%%%%%%%%%%%%%%%%%
\begin{titlepage}
\begin{center}

~
\vskip 1.5cm
{\Large \bf Light neutrinophilic WIMP in the \texorpdfstring{U(1)$_{\rm B-L+xY}$}{} model}

\vskip 2.0cm
{\large
Tatsuya Aonashi$^1$,
Shigeki Matsumoto$^1$, \\ [.5em]
Yu Watanabe$^2$,
and
Yuki Watanabe$^1$
}

\vskip 1.5cm
$^1${\sl Kavli IPMU (WPI), UTIAS, University of Tokyo, Kashiwa, 277-8583, Japan} \\ [.3em]
$^2${\sl Department of Physics and Astronomy, University of California, Los Angeles, California, 90095-1547, USA}\\ [.3em]

\vskip 3.5cm
\begin{abstract}
\noindent
Sub-GeV dark matter is an appealing thermal target because it can still be produced via the standard freeze-out mechanism; at such low masses, achieving freeze-out naturally points to the presence of a light mediator, which shifts the most promising discovery avenues from the energy frontier to the intensity frontier. Realizing this picture is nonetheless challenging, since CMB observations tightly constrain energy injection from dark-matter annihilation at recombination and therefore strongly disfavor simple $s$-wave annihilation into visible Standard-Model final states. In this work, we propose a concrete neutrinophilic framework for sub-GeV thermal dark matter (``light WIMPs'') based on an additional gauge symmetry $\mathrm{U}(1)_{\mathrm{B}-\mathrm{L}+x\mathrm{Y}}$; for an appropriate choice of $x$, the new gauge boson couples predominantly to dark matter and neutrinos while its couplings to charged leptons are suppressed, so that sub-GeV dark matter annihilates almost exclusively into neutrinos, with hadronic modes kinematically closed. We map the parameter space in which the observed relic abundance is reproduced via standard thermal freeze-out in a conventional cosmological history, and show that sizable regions remain viable after imposing current cosmological, indirect-detection, and terrestrial constraints; in part of the allowed parameter space, the dark matter also exhibits sufficiently large self-interactions to potentially alleviate small-scale structure tensions.

%Light (sub-GeV) WIMPs are an appealing target for thermal dark matter because they retain the predictive freeze-out mechanism while shifting discovery prospects toward the intensity frontier and opening connections to light mediators and rich hidden-sector dynamics. Realizing such candidates is, however, challenging: CMB limits on energy injection at recombination tightly constrain $s$-wave annihilation into visible final states, and achieving the correct relic abundance via thermal freeze-out typically requires mediator interactions that are strong enough to maintain equilibrium and drive freeze-out, yet weak enough to evade laboratory, astrophysical, and BBN constraints. In this work, we address these challenges by constructing a concrete neutrinophilic WIMP framework based on an extra gauge symmetry $\mathrm{U}(1)_{\mathrm{B}-\mathrm{L}+x\mathrm{Y}}$, where an appropriate choice of $x$ makes the new gauge boson couple predominantly to neutrinos while strongly suppressing its couplings to charged leptons, so that sub-GeV dark matter annihilates almost exclusively into neutrinos (with hadronic modes kinematically closed). We then delineate the parameter space in which the observed relic abundance is reproduced via standard thermal freeze-out in a conventional cosmology and show that ample viable regions remain after imposing current bounds from cosmology, indirect searches, and terrestrial experiments; moreover, part of the surviving parameter space features sufficiently large dark-matter self-interactions to potentially alleviate small-scale structure tensions.
\end{abstract}

\end{center}
\end{titlepage}

%%%%%%%%%%%%%%%%%%%%%%%%%%%%%%%%
%%%%%%%%%%% Contents %%%%%%%%%%%
%%%%%%%%%%%%%%%%%%%%%%%%%%%%%%%%
\tableofcontents
\newpage
\setcounter{page}{1}

%%%%%%%%%%%%%%%%%%%%%%%%%%%%%%%%%%%%
%%%%%%%%%%% Introduction %%%%%%%%%%%
\section{Introduction}
\label{sec: introcution}

Dark matter is a major open problem at the intersection of particle physics, cosmology, and astrophysics. Theory and observations allow its mass to span an enormous range, from $10^{-22}\,\mathrm{eV}$ to $10^{35}\,\mathrm{g}$, motivating many candidates\,\cite{Preskill:1982, Dodelson:1994, Redondo:2009, Chapline:1975, Bertone:2005, Feng:2010}. A particularly compelling class is {\bf thermal dark matter}: it was once in thermal equilibrium with the Standard Model (SM) plasma in the early Universe and later decoupled via the freeze-out mechanism (or a variant)\,\cite{Lee:1977, Gondolo:1990dk}. This framework is attractive because the relic abundance is set dynamically, the same interactions can be tested experimentally today, and the freeze-out paradigm is an integral part of the standard cosmological framework that has successfully described the thermal history of the Universe. In addition, thermal dark matter often arises in well-motivated SM extensions, such as models of electroweak symmetry breaking, grand unification, and neutrino mass generation\,\cite{Goldberg:1983, Ellis:1984, Ma:2006}. Basic cosmology also implies meaningful mass bounds: Big Bang nucleosynthesis (BBN) suggests that a thermal relic should be heavier than roughly the MeV scale so as not to spoil its successful predictions\,\cite{Boehm:2013jpa}, while requiring freeze-out to occur when the Universe is much colder than the dark matter mass, together with unitarity, yields an upper bound on the mass of a thermal relic\,\cite{Griest:1990}. These arguments motivate the commonly quoted viable window from $\mathcal{O}(1)\,\mathrm{MeV}$ up to $\mathcal{O}(100)\,\mathrm{TeV}$. Within it, electroweak-scale candidates with masses around $1~\mathrm{GeV}$--$1~\mathrm{TeV}$ are known as weakly interacting massive particles (WIMPs) and have been studied extensively, partly because they arise naturally in well-motivated SM extensions such as supersymmetry and other frameworks tied to electroweak symmetry breaking\,\cite{Jungman:1996}. Despite an extensive experimental program, no robust signal has been found, which has shifted attention to thermal dark matter outside the traditional WIMP window. In this article, we focus on the light end, below the GeV scale, often called {\bf sub-GeV dark matter} or {\bf light WIMPs} (we use the latter terminology below).

Cosmological observations place strong constraints on light WIMPs: Cosmic Microwave Background (CMB) anisotropies limit the energy injected into the visible plasma by WIMP annihilation at recombination, implying a robust upper bound on the annihilation rate at that epoch\,\cite{Slatyer:2016, Kawasaki:2021}. For simple $s$-wave annihilation with an essentially velocity-independent cross-section, this bound is in tension with standard freeze-out for sub-GeV WIMPs, since the cross-section allowed at recombination is too small to reproduce the observed relic abundance. Proposed ways to evade this include annihilation mainly into ``invisible'' final states such as neutrinos\,\cite{Herms:2023, Drees:2022}, velocity-dependent annihilation that is efficient at freeze-out but suppressed at recombination\,\cite{B_hm:2004, Bernreuther_2021}, alternative relic-setting mechanisms such as coannihilation or $3\to2$ dynamics\,\cite{Griest:1991, Hochberg:2014, Kuflik:2016}, or freeze-out in a non-standard cosmological history that modifies the relation between the annihilation rate and the final density\,\cite{Kamionkowski:1990, Kane:2015}. In this article, we focus on the first possibility, namely light WIMPs whose dominant annihilation channels are into neutrinos(i.e., {\bf neutrinophilic WIMPs}). This option, however, is not automatic: because neutrinos and left-handed charged leptons form the same $\mathrm{SU}(2)_L$ doublet, interactions with neutrinos typically induce comparable couplings to charged leptons, making it challenging to realize neutrinophilic WIMPs in a field-theoretically consistent way. Meanwhile, collider limits further motivate taking the light WIMP to be an SM singlet; one usually introduces an additional mediator (also an SM singlet) to connect the dark sector to the SM while ensuring DM stability and renormalizable interactions\,\cite{essig2013darksectorsnewlight, alexander2016darksectors2016workshop}. This suggests a simple picture in which the WIMP couples to a new mediator that in turn couples mainly to neutrinos. Representative realizations include a vector mediator coupled to right-handed neutrinos, which mix with active neutrinos and thereby provide an effective neutrino portal\,\cite{Bell_2025, Abdallah:2021NeutrinophilicZp}, and a flavor-dependent gauge boson such as $\mathrm{U}(1)_{L_\mu-L_\tau}$, which couples only to the second and third lepton generations and can kinematically forbid charged-lepton final states in the light-mass regime\,\cite{Drees:2022}. Finally, explicit neutrinophilic models should also be consistent with neutrino-mass generation and possibly baryogenesis, but the simplest versions of these representatives are often not compatible with minimal seesaw and standard leptogenesis without additional fields or more elaborate model building\,\cite{Yanagida:1979as, Fukugita:1986hr, Buchmuller:2005eh}.

In this article, we propose a new neutrinophilic WIMP scenario in which dark matter annihilates predominantly into neutrinos. We extend the SM by an extra Abelian gauge symmetry $\mathrm{U}(1)_{\mathrm{B}-\mathrm{L}+x\mathrm{Y}}$, where $\mathrm{B}$, $\mathrm{L}$, and $\mathrm{Y}$ denote baryon number, lepton number, and hypercharge, and $x$ is a real parameter; for $x \simeq -1/\cos^2{\theta_{W}}$, the associated gauge boson couples mainly to neutrinos while its couplings to both left- and right-handed charged leptons are strongly suppressed well below the electroweak scale, especially in the sub-GeV regime.\footnote{
    While finalizing this manuscript, we became aware of Ref.\,\cite{esseili2025hidinglightvectorboson}, which discusses a charge-phobic dark photon that is essentially identical to our mediator. Their $\mathrm{U}(1)_{\rm B-L}$ setup, with an appropriate choice of kinetic mixing, is related to our $\mathrm{U}(1)_{\rm B-L+xY}$ description for a suitable choice of $x$, as summarized in Appendix~\ref{app: kinetic mixing}.}
In the sub-GeV mass range, annihilation into hadrons is kinematically inaccessible, so the WIMP can annihilate essentially only into neutrinos through this mediator. We determine the parameter region where the observed relic abundance is reproduced via standard thermal freeze-out in a conventional cosmological history, and we show that ample viable space remains consistent with current bounds from cosmology, astrophysical indirect searches, and terrestrial probes (including underground experiments and accelerator searches). Moreover, we demonstrate that part of the allowed region realizes self-interacting dark matter, with self-scattering strong enough to help alleviate small-scale structure tensions. This article is organized as follows: Sec.\,\ref{sec: scenario} introduces the $\mathrm{U}(1)_{\mathrm{B}-\mathrm{L}+x\mathrm{Y}}$ setup and the key quantities used throughout; Sec.\,\ref{sec: conditions and constraints} summarizes the phenomenology, including freeze-out production, the self-interaction requirement, and existing constraints; Sec.\,\ref{sec: results} presents our main results and identifies the allowed parameter space that explains the relic abundance via conventional freeze-out while highlighting the self-interacting region; and Sec.\,\ref{sec: summary} concludes.

%%%%%%%%%%%%%%%%%%%%%%%%%%%%%%%%
%%%%%%%%%%% Scenario %%%%%%%%%%%
%%%%%%%%%%%%%%%%%%%%%%%%%%%%%%%%
\section{The neutrinophilic WIMP}
\label{sec: scenario}

We discuss a light neutrinophilic WIMP (Weakly Interacting Massive Particle) as a dark matter candidate, particularly focusing on a scalar WIMP realized in the gauged U(1)$_{\rm B-L+xY}$ extension of the Standard Model (SM), where ${\rm x} \simeq -1/\cos^2\theta_W$. Here, B, L, and Y denote the baryon number, lepton number, and hypercharge, respectively, and $\theta_W$ is the Weinberg angle. The WIMP is assumed to be charged under this U(1)$_{\rm B-L+xY}$ symmetry. The minimal Lagrangian describing the WIMP and its interactions with SM particles is given by
\begin{align}
    &\mathscr{L} =
      \mathcal{L}_{\rm SM}
    + \mathcal{L}_{\rm BLxY}
    + \mathcal{L}_{\rm DM},
    \nonumber \\
    &\mathcal{L}_{\rm BLxY} =
    -\frac{1}{4} V^{\mu\nu} V_{\mu\nu}
    +|D_\mu S|^2
    +\sum_{i = 1}^3 \bar{N}_i\,i \gamma^\mu D_\mu N_i
    + V(S, H)
    \nonumber \\
    & \qquad~~~
    -\tilde{g}\,V_\mu\,J_{\rm B-L+xY}^\mu
    -\sum_{i, j = 1}^3
    \left[
        y_{ij}^{(\nu)} \bar{L}_i H N_j
        +\frac{1}{2} y_{ij}^{(N)} \bar{N}_i^c N_j S
        + h.c.
    \right],
    \nonumber \\
    &\mathcal{L}_{\rm DM} =
    |D_\mu \varphi|^2 - M_\varphi^2 |\varphi|^2
    -\frac{\lambda_{\varphi H}}{4} |\varphi|^2 |H|^2
    -\frac{\lambda_{\varphi S}}{4} |\varphi|^2 |S|^2
    - \frac{\lambda_\varphi}{4} |\varphi|^4,
    \label{eq: lagrangian 1}
\end{align}
where $\mathcal{L}_{\rm SM}$ denotes SM Lagrangian; $V_\mu$ is the gauge boson associated with the U(1)$_{\rm B-L+xY}$ symmetry; $S$ is the Abelian "Higgs" field that spontaneously breaks the U(1)$_{\rm B-L+xY}$ symmetry; $N_i$ represents the right-handed neutrino (where "$i$" is the generation index) introduced to ensure the model is anomaly-free; $L_i$ denotes the SM lepton doublet; $H$ is the SM Higgs doublet; and $\varphi$ is the scalar WIMP.\footnote{
    We omit the kinetic mixing term between the hypercharge and U(1)$_{\rm B-L+xY}$ gauge bosons in the Lagrangian, assuming it is negligibly small. Including this term does not affect our discussion, as examined in appendix\,\ref{app: kinetic mixing}.}
The quantum numbers of the new particles are summarized in Table\,\ref{tab: representations}. A $Z_2$ symmetry is imposed to ensure the stability of the WIMP, under which all other particles are even. The covariant derivative acting on a particle $f$ is given by  
\begin{equation}
    D_\mu = \partial_\mu + i g_s T^a_G G_\mu^a + i g T^a_W W_\mu^a + i g' Y_f B_\mu + i \tilde{g} q_f V_\mu,
\end{equation}
where $G_\mu^a$, $W_\mu^a$, and $B_\mu$ are the gauge fields of SU(3)$_c$, SU(2)$_L$, and U(1)$_Y$ groups in the SM. The generators of the SU(3)$_c$ and SU(2)$_L$ groups are denoted by $T^a_G$ and $T^a_W$, while $Y_f$ and $q_f$ represent the hypercharge and the B-L+xY charge of the particle $f$. The corresponding gauge couplings of the gauge interactions are denoted by $g_s$, $g$, $g'$, and $\tilde{g}$, respectively.

\begin{table}[t]
    \centering
    \begin{tabular}{c|ccccc}
        & SU(3)$_c$ & SU(2)$_L$ & U(1)$_Y$ & U(1)$_{\rm B-L+xY}$ & $Z_2$ \\
        \hline
        $S$ & {\bf 1} & {\bf 1} & 0 & 2 & + \\
        $N_i$ & {\bf 1} & {\bf 1} & 0 & $-1$ & $+$ \\
        $\varphi$ & {\bf 1} & {\bf 1} & 0 & $q_\varphi$ & $-$ \\
        \hline
    \end{tabular}
    \caption{\small\sl Quantum numbers of the new particles, with $x \simeq -1/\cos^2\theta_W$.}
    \label{tab: representations}
\end{table}

The scalar potential involving the Abelian Higgs field $S$ and the SM Higgs doublet field $H$ is denoted by $V(S, H)$ in Eq.\,(\ref{eq: lagrangian 1}); we omit its explicit form as it is irrelevant to the following discussion. The B-L+xY current of SM fermions for the interaction $\tilde{g}\,V_\mu\,J_{\rm B-L+xY}^\mu$ is give by
\begin{align}
    J_{\rm B-L+xY}^\mu = \sum_f q_f \bar{f} \gamma^\mu f,
\end{align}
where $f = Q_i, U_i, D_i, L_i$, and $E_i$; these represent quark doublet, up-type quark singlet, down-type quark singlet, lepton doublet, and charged lepton singlet of the SM, respectively. The generation index is denoted by "$i$", and each field has a U(1)$_{\rm B-L+xY}$ charge $q_f$, given by $q_{Q_i} \simeq 1/3-1/(6c_W^2)$, $q_{U_i} \simeq 1/3-2/(3c_W^2)$, $q_{D_i} \simeq 1/3+1/(3c_W^2)$ and $q_{L_i} \simeq -1+1/(2c_W^2)$, $q_{E_i} \simeq -1+1/c_W^2$, respectively, where $c_W = \cos \theta_W$. The terms in the square brackets represent the Yukawa interactions of neutrinos $\nu_i \subset L_i$ and $N_i$, which are also irrelevant to the following discussion. The U(1)$_{\rm B-L+xY}$ gauge interaction of the WIMP arises from its kinetic term,
\begin{align}
    \mathcal{L}_{\rm DM} \supset
    |D_\mu\varphi|^2 =
    |\partial_\mu\varphi|^2
    - \tilde{g}\,V^\mu\,q_\varphi\,
    i\,(\varphi^* \overleftrightarrow{\partial_\mu} \varphi)
    +
    \tilde{g}^2\,q_\varphi^2\,V^\mu V_\mu |\varphi|^2.
\end{align}

In this article, we consider the case where the scalar WIMP and the U(1)$_{\rm B-L+xY}$ gauge boson have masses of $\mathcal{O}(1$--$10)\,\mathrm{MeV}$, while other new particles, namely, the right-handed neutrinos and the Abelian Higgs boson, are much heavier than the electroweak scale. Thus, the traditional seesaw\,\cite{Yanagida:1979as, Gell-Mann:1979vob} and leptogenesis\,\cite{Fukugita:1986hr,Buchmuller:2005eh} mechanisms remain valid in this scenario for explaining the masses and mixing of the active neutrinos, as well as the baryon asymmetry of the universe.\footnote{
    The model discussed here is equivalent to the $U(1)_{\rm B-L}$ extension of the SM with an appropriate kinetic mixing term when $\tilde{g} \ll 1$, as shown in appendix\,\ref{app: kinetic mixing}. Therefore, the same mechanisms apply in this framework.}
Assuming such a mass spectrum, the Lagrangian\,(\ref{eq: lagrangian 1}) leads to the effective Lagrangian that describes the physics below the electroweak scale as follows:
\begin{align}
    \mathscr{L}_{\rm EW} =
    \mathcal{L}_{\rm SM}
    -\frac{1}{4} V^{\mu\nu} V_{\mu\nu}
    + \frac{\tilde{m}_{Z'}^2}{2} V^\mu V_\mu
    -\tilde{g} V_\mu\,J_{\rm B-L+xY}^\mu
    +
    \mathcal{L}_{\rm DM}
    +\cdots,
    \label{eq: lagrangian 2}
\end{align}
where the vector field $V_\mu$ acquires a mass term given by $\tilde{m}_{Z'} \equiv 2 \tilde{g} v_S$, through the Higgs mechanism associated with the spontaneous breaking of $\mathrm{U}(1)_{\rm B-L+xY}$, where $v_S$ is the vacuum expectation value of the Abelian Higgs field, i.e., $\langle S \rangle = v_S/\sqrt{2}$.\footnote{
    We consider a very small coupling of $\tilde{g}$, so that $v_S$ can be large enough for leptogenesis to work\,\cite{Choi:2020kch, Lin:2022xbu}.}
We do not explicitly write interactions that are suppressed by the heavy new particles. After electroweak symmetry breaking, the neutral gauge bosons $W^3_\mu$, $B_\mu$, and $V_\mu$ mix with each other. In the unitary gauge, the mass matrix of the bosons, derived from the above Lagrangian, is given by
\begin{align}
    &{\cal L}_{\rm EW} \supset
    \frac{1}{2} (W^{3\mu},\,B^\mu,\,V^\mu)
    \begin{pmatrix}
        g^2 v_H^2/4 & -g g' v_H^2/4 & g \tilde{g} v_H^2/(4 c_W^2) \\
        -g g' v_H^2/4 &  g^{\prime 2} v_H^2/4 & -g' \tilde{g} v_H^2/(4 c_W^2) \\
        g \tilde{g} v_H^2/(4 c_W^2) & -g' \tilde{g} v_H^2/(4 c_W^2) & m_{Z'}^2 
    \end{pmatrix}
    \begin{pmatrix} 
        W^3_\mu \\ B_\mu \\ V_\mu
    \end{pmatrix},
\end{align}
where $H = (0, v_H/\sqrt{2})^T$ with $v_H \simeq 246$\,GeV being the vacuum expectation value of $H$, and $m_Z^2 = \tilde{m}_Z^2 + \tilde{g}^2 v_H^2/4$. The diagonalization of the mass matrix yields mass eigenstates as
\begin{align}
    &
    O {\cal M} O^T \simeq {\rm diag}(m_Z^2, 0, m_{Z'}^2),
    \qquad
    (Z_\mu, A_\mu, Z'_\mu)^T = O\,(W^3_\mu, B_\mu, V_\mu)^T,
    \nonumber \\
    &
    O \simeq
    \begin{pmatrix}
        c_W & -s_W & \displaystyle\frac{\tilde{g}}{g' c_W^2} \frac{s_W m_Z^2}{m_Z^2 - m_{Z'}^2} \\
        s_W & c_W & 0 \\
        \displaystyle -\frac{\tilde{g}}{g' c_W^2} \frac{c_W s_W m_Z^2}{m_Z^2 - m_{Z'}^2} & \displaystyle \frac{\tilde{g}}{g' c_W^2} \frac{s_W^2 m_Z^2}{m_Z^2 - m_{Z'}^2} & 1
    \end{pmatrix},
\end{align}
where $m_Z = (g^2 + g^{\prime 2})^{1/2}\,v_H/2$ represents the mass of the $Z$ boson, and $s_W = \sin\theta_W$. We assume $\tilde{g} \ll 1$ and neglect terms smaller than $\mathcal{O}(\tilde{g})$ to derive the above result. As a consequence, the coupling of a SM fermion ``$f$'' with the vector mediator $Z'$ is given by
\begin{align}
    {\cal L}_{\rm EW} \supset
    -g_f \bar{f}\gamma^\mu f\,Z'_\mu,
    \qquad
    g_f \simeq q_f\,\tilde{g}
    +Q_f \frac{\tilde{g}}{c_W^2}\frac{m_Z^2\,s_W^2}{m_Z^2 - m_{Z'}^2}
    -T_f \frac{\tilde{g}}{c_W^2}\frac{m_Z^2}{m_Z^2 - m_{Z'}^2},
\end{align}
%\begin{align}
%    {\cal L}_{\rm EW} \supset
%    -g_f \bar{f}\gamma^\mu f\,Z'_\mu,
%    \qquad
%    g_f \simeq q^{\rm B-L}_f\,\tilde{g}
%    -Q_f \tilde{g}\frac{m_Z^2 - m_{Z'}^2/c_W^2}{m_Z^2 - m_{Z'}^2}
%    -T_f\tilde{g}\frac{m_{Z'}^2/c_W^2}{m_Z^2 - m_{Z'}^2},
%\end{align}
where $Q_f$ and $T_f$ represent the electric and weak charges of the particle ``$f$'', respectively.

The effective Lagrangian describing the physics well below the GeV scale is obtained from Lagrangian\,(\ref{eq: lagrangian 2}) by integrating out the fields with masses above this scale as follows:
\begin{align}
    {\cal L}_{\rm GeV}
    =
    & {\cal L}_{\rm kin}[A_\mu, Z'_\mu, G_\mu^a, \nu_{L, i}, e, \mu, u, d, s]
    +{\cal L}_{\rm QED}[A_\mu, e, \mu, u, d, s]
    +{\cal L}_{\rm QCD}[G_\mu^a, u, d, s]
    \nonumber \\
    &- i\,g_\varphi\,Z^{\prime \mu}\,
    (\varphi^* \overleftrightarrow{\partial_\mu} \varphi)
    + g_\varphi^2 Z^{\prime \mu} Z'_\mu |\varphi|^2
    - \frac{\lambda_\varphi}{4} |\varphi|^4
    \nonumber \\
    & +\tilde{g} \sum_i\bar{\nu}_{L,i} \slashed{Z}' \nu_{L,i}
    +\frac{\tilde{g}}{3} \bar{u} \slashed{Z}' u
    -\frac{2\tilde{g}}{3} \bar{d} \slashed{Z}' d
    -\frac{2\tilde{g}}{3} \bar{s} \slashed{Z}' s
    + \cdots,
    \label{eq: lagrangian 3}
\end{align}
where $\nu_{L, i}$, $e$, $\mu$, $u$, $d$, and $s$ represent the fields of the neutrino (with a generation index $i$), electron, muon, up quark, down quark, and strange quark, respectively. Their QED and QCD interactions are contained in ${\cal L}_{\rm QED}$ and ${\cal L}_{\rm QCD}$, respectively, while the kinetic terms of all light particles are included in ${\cal L}_{\rm kin}$. We omit explicit expressions for interactions that are suppressed by particles heavier than the GeV scale. It is found that the WIMP $\varphi$ couples to the vector mediator $Z'$, and the mediator couples to neutrinos $\nu_{L, i}$ in a generation-blind manner, as well as to quarks, but not to charged leptons ($e$ and $\mu$), as expected.\footnote{
    Note that the mediator couples to three photons at the one-loop level. However, its coupling is highly suppressed, even compared to the dark photon case\,\cite{McDermott:2017qcg}, as the mediator does not couple to charged leptons.}

The interactions between the vector mediator $Z'$ and the light quarks $u$, $d$, and $s$ should be expressed in terms of hadronic degrees of freedom at a certain low-energy scale. At energies well below the GeV scale, the interactions between the mediator and the lightest mesons (i.e., pions) can be derived using chiral perturbation theory\,\cite{Scherer:2002tk}, as follows:
{\small
\begin{align}
    \mathcal{L}_{\rm GeV} \supset
    &-i \tilde{g} Z^{\prime \mu}\,(\pi^+ \overleftrightarrow{\partial_\mu} \pi^-)
    +\tilde{g}^2\,Z^{\prime \mu}\,Z'_\mu \pi^+  \pi^-
    +2 e \tilde{g}\, A^\mu\, Z'_\mu \pi^+  \pi^-
    -\tilde{g}^2 \frac{\epsilon^{\mu\nu\rho\sigma}}{8\pi^2 f_\pi} (\partial_\mu \pi^0)\,(\partial_\nu\, Z'_\rho)\,Z'_\sigma,
\end{align}
}where $e$ represents the QED coupling, while $\pi^\pm$ and $\pi^0$ denote the charged and neutral pion fields, respectively. The last interaction arises from the WZW term. It is interesting to see that the $\pi^0$-$A$-$Z'$ vertex, which is generally predicted in the WZA term, vanishes due to the B-L+xY charge assignment of the quarks\,(\ref{eq: lagrangian 3}). Furthermore, interactions between the vector mediator and nucleons are obtained using the conservation of the vector currents\,\cite{Hisano:2017jmz, Cirelli:2013ufw}:
\begin{align}
    \mathcal{L}_{\rm GeV} \supset
    -\tilde{g}\,\bar{n} \gamma_\mu \slashed{Z}' n + \cdots,
\end{align}
where we assume that the momenta injected into the vertex are small enough compared to the GeV scale and omit the explicit expression of the terms at the next-to-leading order, such as dipole interactions. It is interesting to see that the vector interaction between the vector mediator $Z'$ and the proton vanishes again due to the B-L+xY charge assignment. 

%\begin{align}
%\mathcal{L}_{\rm GeV} \supset
%    &-i(\xi g' \cos^2\theta_W)\,Z^{\prime \mu}\,(\pi^+ \overleftrightarrow{\partial_\mu} \pi^-) 
%    -(\xi g' \cos^2\theta_W)^2\,Z^{\prime \mu}\,Z'_\mu\pi^+\pi^-
%    \nonumber \\
%    &+2e(\xi g' \cos^2\theta_W)\,A^{\prime \mu}\,Z'_\mu\pi^+\pi^-
%    \nonumber\\
%    &-\frac{\epsilon^{\mu\nu\rho\sigma}}{8\pi^2 f_\pi} (\partial_\mu \pi^0)
%    \left[
%        e\,(g_{\rm B-L} - \xi g' \cos^2\theta_W)\left\{ (\partial_\nu\, A_\rho)\,Z'_\sigma + (\partial_\nu\, Z'_\rho)\,A_\sigma \right\}
%    \right.
%    \nonumber\\
%    &\left.
%        -(\xi g' \cos^2\theta_W)\,(2g_{\rm B-L} - \xi g' \cos^2\theta_W)\, (\partial_\nu\, Z'_\rho)\, Z'_\sigma
%    \right]
%\end{align}

As a result, when the vector mediator $Z'$ is lighter than twice the pion mass, it always decays invisibly (i.e., into pairs of neutrinos and WIMPs), with the branching fractions of
\begin{align}
    &\Gamma(Z' \to \nu_i \bar{\nu}_i) =
    \frac{\tilde{g}^2}{24\pi} m_{Z'}
    \nonumber \\
    &\Gamma(Z' \to \varphi \varphi^*) =
    \frac{g_\varphi^2}{48\pi} m_{Z'}
    \left( 1 - 4 m_\varphi^2/m_{Z'}^2 \right)^{3/2},
    \label{eq: decay width}
\end{align}
with $m_\varphi$ being the physical mass of the scalar WIMP. Although such an invisible mediator is usually probed efficiently in charged-lepton collider and beam-dump experiments\,\cite{Graham:2021ggy, Ferber:2023iso}, the mediator predicted in the model cannot be efficiently explored in such experiments, as it does not couple to charged leptons (nor to protons at leading order). Consequently, the $Z'$ boson should instead be produced and searched for in rare decays of heavy mesons at high-luminosity experiments, as well as in collisions involving quarks at high-energy facilities.

Meanwhile, if the WIMP is lighter than the charged pion (i.e., $m_\varphi < m_{\pi^\pm} \simeq 139$\,MeV), its dominant annihilation process into SM particles is $\varphi \varphi^* \to \nu_i \bar{\nu}_i$. The annihilation cross-section, multiplied by the relative velocity between the incident WIMPs, is given by
\begin{align}
    \sigma v (\varphi \varphi^* \to \nu_i \bar{\nu}_i)
    &=
    \frac{\tilde{g}^2\,g_\varphi^2}{12\pi}
    \frac{s - 4 m_{\varphi}^{2}}
    {(s - m_{Z'}^2)^2 + s\,\Gamma_{Z'}^2(s)},
    \label{eq: Ann Xsection}
\end{align}
where $\Gamma_{Z'}(s) \equiv [3\,\Gamma(Z' \to \nu_i \bar{\nu}_i) + \Gamma(Z' \to \varphi \varphi^*)]_{m_{Z'} \to \sqrt{s}}$ denotes the total width appearing in the $Z'$ propagator, with $s$ being the center-of-mass energy. 
The WIMP can also annihilate into a pair of mediators ($Z'$s) when $m_\varphi > m_{Z'}$, and each $Z'$ produced in this process subsequently decays into a neutrino pair. 
The corresponding annihilation cross-section is given by
\begin{align}
    \sigma v (\varphi \varphi^* \to Z' Z')
    &\simeq
    \frac{4 g_\varphi^4\,
    (3 m_{Z'}^4 - 8 m_{Z'}^2 m_\varphi^2 + 8 m_\varphi^4) (m_\varphi^2-m_{Z'}^2)^{1/2}}{128 \pi m_\varphi^3\,(2m_\varphi^2 - m_{Z'}^2)^2},
\end{align}
Here we consider the non-relativistic limit of the incident WIMP particles. However, in the numerical analysis presented in the following sections, the annihilation cross-section is evaluated without imposing this limit. Since the WIMP annihilates exclusively into neutrinos in all cases where $m_\varphi < m_{\pi^\pm}$, it is referred to as the neutrinophilic WIMP in this article.

On the other hand, although the WIMP does not couple to protons at leading order, it can still scatter off nuclei through its coupling to neutrons within the nucleus.\footnote{
    The WIMP scatters off a proton via the $Z$ and Higgs bosons; their contributions are negligibly small.}
The scattering cross-section between the WIMP and a neutron is given by the following expression:
\begin{align}
    &\sigma (\varphi^{(*)} n \to \varphi^{(*)} n) =
    \frac{m_n^2 m_\varphi^2\,\tilde{g}^2\,g_\varphi^2}
    {\pi m_{Z'}^4 (m_\varphi + m_n)^2},
    \label{eq: scattering cross-section}
\end{align}
where $m_n$ is the mass of the neutron. In direct detection experiments, the recoil energy from scattering is generally too small to be reliably distinguished from the background for WIMP masses below 100\,MeV\,\cite{Schumann:2019eaa}. Although such light WIMPs are usually expected to be efficiently probed through electron rather than nuclear scattering\,\cite{Essig:2011nj}, the neutrinophilic WIMP considered here remains particularly elusive, as it does not couple to electrons.
Finally, the neutrinophilic scalar WIMP may exhibit a relatively large self-scattering cross-section, particularly when it is light. Its differential self-scattering cross-section is
\begin{align}
    &\frac{d\sigma (\varphi \varphi^* \to \varphi \varphi^*)}{d\Omega} =
    \frac{1}{64 \pi^{2} \,s}
    \left|
        \lambda_\varphi
        +\frac{g_\varphi^2 (t - u)}{s - m_{Z'}^2 + i\sqrt{s}\,\Gamma_{Z'}(s)}
        +\frac{g_\varphi^2 (s -u)}{t - m_{Z'}^2}
    \right|^2,
    \nonumber \\
    &\frac{d\sigma (
    \varphi^{(*)} \varphi^{(*)} \to \varphi^{(*)} \varphi^{(*)})}{d\Omega} =
    \frac{1}{64 \pi^{2} \,s}
    \left|
        \lambda_\varphi
        +\frac{g_\varphi^2 (u - s)}{t - m_{Z'}^2}
        +\frac{g_\varphi^2 (t - s)}{u - m_{Z'}^2}
    \right|^2,
    \label{eq: self-scattering Xsection}
\end{align}
where $t$ and $u$ are the so-called Mandelstam variables, which depend on the center-of-mass energy squared $s$ and the angle between the incoming and outgoing WIMPs. The total scattering cross-section is obtained by integrating the differential cross-section over the angle.

Another specific prediction of the model is the existence of non-standard neutrino interactions. The first example is neutrino self-interactions. Although such interactions exist in the SM via the exchange of weak gauge bosons, neutrinos can also interact with each other through the exchange of $Z'$. This interaction can be significantly stronger than that in the SM when the gauge coupling $\tilde{g}$ is sufficiently large and the mediator mass $m_{Z'}$ is sufficiently small. The self-interaction between neutrinos $\nu_i$ and $\nu_j$ (or anti-neutrinos $\bar{\nu}_i$ and $\bar{\nu}_j$) occurs via $Z'$ exchange in the $t$-channel when $i \neq j$, and via both the $t$- and $u$-channels when $i = j$. In contrast, the interaction between $\nu_i$ and $\bar{\nu}_j$ proceeds via the $t$-channel for $i \neq j$, and via both the $t$- and $s$-channels when $i = j$, with the $s$-channel giving rise to resonant self-scattering when the incoming neutrino energy is comparable to the mediator mass $m_{Z'}$. All of the differential scattering cross-sections are derived from the Lagrangian~(\ref{eq: lagrangian 3}) as
%\begin{align}
%    &
%    \frac{d\sigma(\nu_i \nu_j \to \nu_i \nu_j)}{d\Omega} =
%    \frac{d\sigma(\bar{\nu}_i \bar{\nu}_j \to \bar{\nu}_i \bar{\nu}_j)}{d\Omega} =
%    \frac{1}{2^{\delta_{ij}}}
%    \frac{\tilde{g}^4}{32 \pi^2} s
%    \left| \frac{1}{t-m_{Z'}^2} +\frac{\delta_{ij}}{u-m_{Z'}^2} \right|^2,
%    \nonumber \\
%    &
%    \frac{d\sigma(\nu_i \bar{\nu}_j \to \nu_i \bar{\nu}_j)}{d\Omega} = 
%    \frac{\tilde{g}^4} {32\pi^2} \frac{u^2}{s}
%    \left| \frac{1}{t-m_{Z'}^2} +\frac{\delta_{ij}}{s-m_{Z'}^2+ i\sqrt{s}\,\Gamma_{Z'}(s)} \right|^2. 
%\end{align}
\begin{align}
    &
    \frac{d\sigma(\nu_i \nu_j \to \nu_i \nu_j)}{d\Omega} =
    \frac{1}{2^{\delta_{ij}}}
    \frac{\tilde{g}^4}{32 \pi^2} s
    \left[
    \frac{1 + (u/s)^2}{(t-m_{Z'}^2)^2} +\delta_{ij}\frac{1 + (t/s)^2}{(u-m_{Z'}^2)^2} + \frac{2\delta_{ij}}{(t - m_{Z'}^2)(u - m_{Z'}^2)} 
    \right],
    \\
    &
    \frac{d\sigma(\nu_i \bar{\nu}_j \to \nu_i \bar{\nu}_j)}{d\Omega} = 
    \frac{\tilde{g}^4} {32\pi^2} \frac{u^2}{s}
    \left[
     \frac{1 + (s/u)^2}{(t-m_{Z'}^2)^2} +\delta_{ij}\frac{1 + (t/u)^2}{|s-\tilde{m}_{Z'}^2|^2} + \Re \left\{\frac{2 \delta_{ij}}{(t - m_{Z'}^2)(s-\tilde{m}_{Z'}^2)} \right\}
     \right],
     \nonumber
\end{align}
where $d\sigma(\bar{\nu}_i \bar{\nu}_j \to \bar{\nu}_i \bar{\nu}_j)/d\Omega = d\sigma(\nu_i \nu_j \to \nu_i \nu_j)/d\Omega$ and $\tilde{m}_{Z'}^2 = m_{Z'}^2 - i\,s^
{1/2}\,\Gamma_{Z'}(s)$.

Moreover, neutrinos can also coherently scatter off a nucleus or a dark matter particle via the exchange of the leptophilic mediator $Z'$, in addition to the processes predicted by the SM (i.e., via exchange of weak gauge bosons). These interactions play a crucial role in detecting new physics signals in astrophysical observations, particularly when $\tilde{g}$ is sufficiently large and $m_{Z'}$ is sufficiently small. The corresponding scattering cross-sections are given by
\begin{align}
    &
    \frac{d \sigma(\nu_i n \to \nu_i n)}{d \Omega} = \frac{\tilde{g}^4}{32 \pi^2} \frac{2(s - m_n^2)^2 + 2st + t^2}{(m_n^2 + s)(t - m_{Z'}^2)^2},
    \nonumber \\
    &
    \frac{d \sigma(\nu_i \varphi^{(*)} \to \nu_i \varphi^{(*)})}{d \Omega} = \frac{\tilde{g}^2 g_\varphi^2}{16 \pi^2} \frac{(s - m_\varphi^2)^2 + st}{(s + m_\varphi^2)(t - m_{Z'}^2)^2},
\end{align}
where both the scattering processes arise from $t$-channel exchange of the mediator $Z'$.

%%%%%%%%%%%%%%%%%%%%%%%%%%%%%%
%%%%%%%%%%% Status %%%%%%%%%%%
%%%%%%%%%%%%%%%%%%%%%%%%%%%%%%
\section{Phenomenology of the neutrinophilic WIMP}
\label{sec: conditions and constraints}

In this section, we investigate the phenomenology of the neutrinophilic scalar WIMP model constructed in the previous section. We begin by examining the role of dark matter self-interactions, which have been proposed as a possible solution to the small-scale structure problems in cosmology. We then analyze the conditions under which the model reproduces the observed relic abundance, thereby accounting for the present dark matter density in the Universe. Next, we consider cosmological and astrophysical constraints, including those from the CMB, BBN, and indirect searches for dark matter. Finally, we discuss implications for terrestrial experiments, such as direct detection through dark matter--nucleus scattering, coherent elastic neutrino--nucleus scattering, and searches at $K$-meson factories.

\subsection{Model parameters}

Before proceeding to the discussion of the relevant conditions and constraints, we provide a brief summary of the model parameters that are subject to scanning in order to determine the viable parameter space for the neutrinophilic WIMP model. As specified by the Lagrangian~(\ref{eq: lagrangian 3}), the model is characterized by three independent couplings, $\tilde{g}$, $g_\varphi$, and $\lambda_\varphi$, as well as the masses of the WIMP and the mediator, denoted by $m_\varphi$ and $m_{Z'}$, respectively. Accordingly, the model involves five parameters associated with new physics in total.

If we focus on the resonance parameter region where $m_{Z'} \simeq 2 m_\varphi$, which is required to satisfy the self-scattering condition discussed in the next subsection, we may introduce $v_{\rm R} \equiv 2\, (m_{Z'}/m_\varphi - 2)^{1/2} > 0$ as an alternative to $m_{Z'}$ as an input model parameter. This parameter corresponds to the relative velocity of incident WIMPs that resonate with the $Z'$ propagator. Using $v_{\rm R}$, the annihilation cross-section\,(\ref{eq: Ann Xsection}) and the self-scattering cross-section\,(\ref{eq: self-scattering Xsection}) of the scalar WIMP in the non-relativistic limit are respectively expressed as
\begin{align}
    &\sigma v(\varphi \varphi^* \to \nu_i \bar{\nu}_i) \simeq
    \frac{\tilde{g}^2 g_\varphi^2\,v^2}
    {3 \pi m_{Z'}^2[(v^2 - v_R^2)^2 + 16\gamma_{Z'}^2(v)]},
    \\
    &\sigma (\varphi \varphi^* \to \varphi \varphi^*) \simeq
    2\sigma_0
    +\frac{g_\varphi^4\,v^4}{48 \pi m_{Z'}^2
    [(v^2 - v_R^2)^2 + 16\gamma_{Z'}^2(v)]},
    \\
    &\sigma (\varphi \varphi \to \varphi \varphi) =
    \sigma (\varphi^* \varphi^* \to \varphi^* \varphi^*) \simeq
    \sigma_0,
\end{align}
where $\gamma_{Z'}(v) \simeq (48\,\tilde{g}^2 + g_\varphi^2\,v^3)/(384\pi)$. We also adopt $\sigma_0 \equiv \lambda_{\varphi}^{2}/(128 \pi m_\varphi^2)$ in place of the coupling $\lambda_\varphi$ as an input model parameter. The contributions from the $t$- and $u$-channel diagrams to the scattering cross-sections are neglected, as the condition $\lambda_\varphi \gg g_\varphi^2$ is required by the constraints discussed below. Moreover, the contribution from the interference between the contact and $s$-channel diagrams in the annihilation cross-section is also negligible\,\cite{Chu:2018fzy}.

\subsection{The self-scattering condition}
\label{subsec: self-scattering}

\begin{figure}[t]
    \centering    \includegraphics[keepaspectratio, scale=0.41]{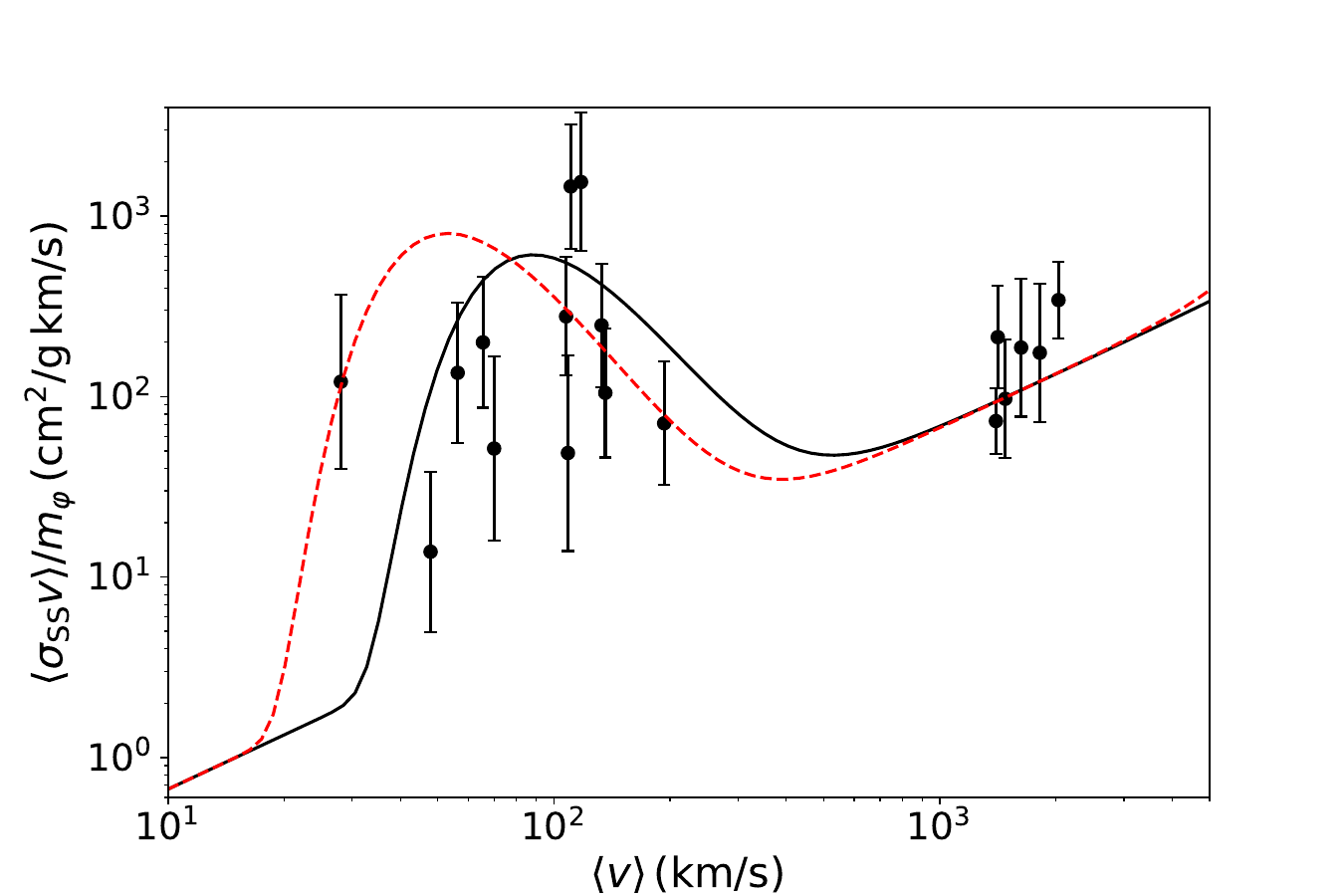}
    \caption{\small \sl The velocity dependence of the self-scattering cross-section (multiplied by the relative velocity between incident particles, $v$) predicted by the neutrinophilic WIMP is shown. The prediction using the parameter set $(m_\varphi, v_R, g_\varphi, \sigma_0/m_\varphi) = (7.7\,\mathrm{MeV}, 108\,\mathrm{km/s}, 2.4 \times 10^{-3}, 0.067\,\mathrm{cm}^2/\mathrm{g})$ is represented by a black solid line, while the one using the parameter set $(7.7\,\mathrm{MeV}, 65\,\mathrm{km/s}, 3.1 \times 10^{-3}, 0.067\,\mathrm{cm}^2/\mathrm{g})$ is represented by a red dashed line. The data obtained from kinematical observations are shown as black dots with error bars\,\cite{Kaplinghat:2015aga}. See the main text for more details.}
    \label{fig: self-scattering}
\end{figure}

The diversity problem of DM density profiles at galactic centers can be addressed if DM exhibits a sufficiently large self-scattering cross-section\,\cite{Spergel:1999mh, Kamada:2016euw}. The magnitude and velocity dependence of the cross-section required to resolve this issue have been quantitatively analyzed in Ref.\,\cite{Kaplinghat:2015aga}, as illustrated in Fig.\,\ref{fig: self-scattering}. The figure presents observational data from five galaxy clusters\,\cite{Newman:2012nw}, seven low-surface-brightness spiral galaxies\,\cite{KuziodeNaray:2007qi}, and six dwarf galaxies from the THINGS sample\,\cite{Oh:2010ea}. In the scenario considered here, where the present-day DM consists equally of $\varphi$ and $\varphi^*$, the effective self-scattering cross-section is given by:
\begin{align}
    \sigma_{\rm SS} =
    \frac{1}{4} \sigma (\varphi \varphi \to \varphi \varphi)
    +\frac{1}{2} \sigma (\varphi \varphi^* \to \varphi \varphi^*)
    +\frac{1}{4} \sigma (\varphi^* \varphi^* \to \varphi^* \varphi^*).
    \label{eq: self-scattering cross-section}
\end{align}
Meanwhile, the bracketed cross-section denoted by $\langle \sigma_{\rm SS} v \rangle_{v_{0,i}}$ on the $y$-axis of the figure corresponds to the velocity-averaged cross-section, computed using the velocity distribution function $f(v; v_{0,i}) \simeq 4\pi^{-1/2} v^2 \exp(-v^2/v_{0,i}^2)/v_{0,i}^3$. Here, $\langle v \rangle_i = 2v_{0,i}/\sqrt{\pi}$ on the $x$-axis denotes the mean relative velocity between incident WIMPs in the $i$-th galaxy or cluster.

When the WIMP mass lies at the MeV scale, the self-scattering condition imposes specific constraints on the model parameters, under the assumption that the mediator ($Z'$) width is predominantly determined by its decay into a pair of WIMPs\,\cite{Chu:2018fzy}.\footnote{
    In the viable parameter region, $Z'$ decays primarily into WIMPs, as discussed in the following section.}
First, a narrow resonance is required, with its width scaling as $\Gamma_{Z'} \propto m_\varphi^4$, in correlation with the WIMP mass. Additionally, the parameters $v_R$ and $\sigma_0/m_\varphi$ are required to be approximately $100\,\mathrm{km/s}$ and $0.1\,\mathrm{cm}^2/\mathrm{g}$, respectively, as indicated by the solid and dashed lines shown in Fig.\,\ref{fig: self-scattering}.

\subsection{The relic abundance condition}

We focus on the scenario in which the present-day DM abundance is explained by the standard thermal freeze-out mechanism, a cornerstone of modern cosmology. In this work, we refer to the requirement that the observed relic density be reproduced as the \textit{relic abundance condition}. This condition is fulfilled when the neutrinophilic WIMP possesses an appropriate annihilation cross-section. Specifically, in the case where $m_\varphi > m_{Z'}$, the WIMP predominantly annihilates into a pair of mediator particles. The relic abundance condition is then satisfied by selecting a suitable value of the coupling $\tilde{g}$ for given values of $m_\varphi$ and $m_{Z'}$. Conversely, when $m_\varphi < m_{Z'}$, the WIMP annihilates into a pair of neutrinos via $s$-channel exchange of the mediator. In this case, the annihilation cross-section depends on the product $g_\varphi \tilde{g}$, and the condition is met by choosing an appropriate value of this product for given masses. Since both couplings are only weakly constrained, as will be discussed below, the annihilation cross-section can reach values as large as $\mathcal{O}(1)$\,pb when both $m_\varphi$ and $m_{Z'}$ are well below 1\,GeV. Therefore, by solving the Boltzmann equation for the WIMP number density in the standard way, we can readily identify viable model parameters that reproduce the observed DM abundance in our universe. In other words, in both regimes discussed above, a wide range of parameter space remains consistent with the relic abundance condition.

On the other hand, if we impose not only the relic abundance condition but also the self-scattering condition (although the latter is not strictly necessary, as the small-scale crisis may be resolved by mechanisms unrelated to dark matter), the model parameters must satisfy nontrivial correlations. Specifically, the relic abundance condition is satisfied by choosing an appropriate value of the coupling $\tilde{g}$ for a given $m_\varphi$, provided that the other parameters, $v_{\rm R}$ and $g_\varphi$, are consistent with the self-scattering requirement discussed in Section\,\ref{subsec: self-scattering}. Since the self-scattering condition requires $v_{\rm R} \sim 10^{-3}$, i.e., $m_{Z'}$ must be nearly degenerate with $2m_\varphi$, the calculation of the relic abundance necessitates special care. As inferred from the self-scattering analysis, the annihilation cross-section between $\varphi$ and $\varphi^*$ is significantly enhanced due to the $s$-channel resonance of the $Z'$ mediator. Consequently, $\tilde{g}$ must be highly suppressed to explain the observed dark matter abundance. This suppression renders the scattering of the WIMP off SM particles extremely weak at the freeze-out epoch, causing kinetic decoupling to occur before chemical decoupling. Such early kinetic decoupling is known to alter the predictions of standard thermal relic abundance calculations\,\cite{Binder:2017rgn}.

An interesting feature of the temperature (i.e., time) evolution of the WIMP dark matter abundance, in the parameter region where the self-scattering condition is satisfied, is that it exhibits only a very mild dependence on the coupling $g_\varphi$. For a given WIMP dark matter mass $m_\varphi$, the thermally averaged annihilation cross-section is given by Eq.\,(\ref{eq: Ann Xsection})\,\cite{Gondolo:1990dk}.
\begin{align}
    \langle \sigma v \rangle \equiv
    \frac{2\pi^2}{[4 \pi x^2 K_2(x)]^2} \int^\infty_{2x} dy\,(\sigma v)\,y^3\,(y^2 - 4 x^2)^{1/2}\,K_1(y).
\end{align}
Here, $x = m_\varphi/T$, and $K_i(x)$ denotes the modified Bessel function of order $i$. The annihilation cross-section in the integrand, whose form is given in Eq.\,(\ref{eq: Ann Xsection}), exhibits a delta function-like behavior peaked at $s \sim m_{Z'}^2$. Thus, its thermal average can be approximated as  
\begin{align}
    \langle \sigma v \rangle \simeq
    \frac{6 \pi \tilde{g}^2}{m_\varphi^2}
    \frac{x K_1[2x + x\,v_R^2/2]}{K_2^2(x)}
    +
    \frac{3 \tilde{g}^2 g_\varphi^2 m_\varphi^2}{2 \pi (4 m_\varphi^2 - m_{Z'}^2)^2 x},
    \label{eq: Thermally Averaged Ann Xsection}
\end{align}
unless $x$ is very small (i.e., the temperature is very high). We use the fact that the total decay width of $Z'$ is predominantly determined by its decay into a WIMP pair. Since the first term dominates except in the late universe, the thermally averaged cross-section is primarily governed by $\tilde{g}$ and is only weakly dependent on the other coupling, $g_\varphi$.

\subsection{Constraints from cosmology}
\label{subsec: cosmological constraints}

A light WIMP with a mass at the MeV scale is expected to have been in thermal equilibrium with SM particles during the epoch of Big Bang Nucleosynthesis (BBN), neutrino decoupling, or in nearby periods. Such a WIMP can affect the evolution of photon and neutrino temperatures, as well as the expansion rate of the universe during and after these epochs\,\cite{Dolgov:2002wy, Ibe:2018juk}. Even after departing from equilibrium, the WIMP continues to annihilate and inject energy into the cosmic plasma composed of SM particles. Additionally, the mediator particle, assumed to be comparably light, can also influence cosmology both directly, by coexisting with the WIMP in the early universe, and indirectly, by inducing a neutrino self-interaction, etc. These effects are subject to constraints, as no signatures of such new physics have been observed in Cosmic Microwave Background (CMB) and BBN measurements\,\cite{Boehm:2013jpa, Giovanetti:2021izc, Sabti:2021reh, Chu:2022xuh}.

In the neutrinophilic scenario considered in this study, the entropy initially carried by the new particles, namely, the WIMP and the mediator, is transferred to neutrinos once these particles become non-relativistic. This entropy transfer alters the evolution of the neutrino temperature, raising it and thereby increasing the expansion rate of the universe. As a result, precise CMB observations impose a stricter constraint on the neutrinophlic scenario than those derived from BBN measurements. The CMB constraint is typically expressed in terms of the so-called effective number of relativistic degrees of freedom, $N_{\rm eff}$, as follows:
\begin{align}
    \rho_R \simeq
    \frac{\pi^2}{15}
    \left[
        1 + \frac{7}{8}
        \left( \frac{4}{11} \right)^{4/3} N_{\rm eff}
    \right] T_\gamma^4,
    %N_{\rm eff} =
    %3 \left[ \frac{11}{4}\left( \frac{T_\nu}{T_\gamma}\right)^3 \right]^{4/3},
\end{align}
where $\rho_R$ and $T_\gamma$ denote the energy density of radiation and the temperature of the electromagnetic sector at the time of recombination, respectively. The CMB observation gives $N_{\rm eff} = 2.99^{+0.34}_{-0.33}$ at the 95\,\% confidence level\,\cite{Planck:2018vyg}. On the other hand, it is well known that the value of the Hubble constant measured from galaxy surveys significantly deviates from that inferred from CMB observations\,\cite{Verde:2019ivm}. This discrepancy may be explained by some new physics beyond the SM, and the corresponding estimate of the effective number of neutrino species, taking this possibility into account, is $N_{\rm eff} = 3.27 \pm 0.30$ at the 95\,\% confidence level\,\cite{ParticleDataGroup:2024cfk}. Both results for $N_{\rm eff}$ are consistent with the SM prediction, namely the case without any additional contributions from the WIMP and the mediator\,\cite{deSalas:2016ztq, Bennett:2020zkv}. Hence, we adopt the latter value in our analysis to take a conservative and robust approach.

The effective number of degrees of freedom, including the contributions from the WIMP and the mediator, is predicted based on the conservation of entropy in the early universe.
\begin{align}
    N_{\rm eff} \simeq
    3
    \left\{
        1 + \frac{30}{7 \pi^2 T_D^3}[s_\varphi(T_D) + s_{Z'}(T_D)]
    \right\}^{4/3}.
    \label{eq: neutrinophilic Neff}
\end{align}
The entropy densities are $s_\varphi(T_D) = m_\varphi^4/(3 \pi^2 T_D)\int^\infty_1 dy (4y^2 - 1)\sqrt{y^2 -1}/(e^{m_\varphi y/T_D} - 1)$ and $s_{Z'}(T_D) = m_{Z'}^4/(2 \pi^2 T_D) \int^\infty_1 dy (4y^2 - 1)\sqrt{y^2 -1}/(e^{m_{Z'} y/T_D} - 1)$, assuming that the new particles are in equilibrium with SM particles (neutrinos) and that neutrinos instantaneously decouple from the thermal bath (electrons, positrons, and photons) when the temperature of the universe reaches $T_D = 1.7\,\text{MeV}\,$\cite{Matsumoto:2018acr, Ibe:2018juk}. When $m_{Z'} \gg m_\varphi$ ($m_{Z'} \simeq 2 m_\varphi$), comparing the resulting effective number $N_{\rm eff}$ with the observational data presented above yields a lower bound on the WIMP mass: $m_\varphi \geq 5.3\,\text{MeV}$ ($5.8\,\text{MeV}$) at the 95\% confidence level. On the other hand, when $m_{Z'} \ll m_\varphi$, the lower bound on $m_{Z'}$ is obtained as $m_{Z'} \geq 6.5\,\text{MeV}$.

Meanwhile, the annihilation cross-section of a light WIMP is generally constrained by both CMB and BBN observations. Even after freeze-out, the WIMP continues to inject energy into the cosmic medium, potentially destroying light elements and altering the recombination history of the universe. In particular, the cross-section is tightly constrained if the WIMP annihilates into electromagnetically interacting particles ($e^\pm$ and $\gamma$), as these can induce electromagnetic showers and produce large numbers of energetic photons. Particles emitted after BBN can destroy light elements through photodisintegration processes\,\cite{Kawasaki:1994af, Ellis:1990nb}, while those emitted during the recombination epoch can enhance the residual ionization. The light neutrinophilic WIMP considered in this article, fortunately, annihilates only into neutrinos when its mass is below the pion mass. As a result, it does not affect either the standard BBN and CMB scenarios. Therefore, in our analysis, we restrict ourselves to the parameter region where the WIMP mass is below the pion mass (i.e., $m_\varphi \lesssim 139\,\mathrm{MeV}$).

On the other hand, the neutrinophilic mediator $Z'$ induces non-standard neutrino interactions, such as neutrino self-interactions and interactions with WIMPs, which could potentially affect BBN, the CMB, and the formation of large-scale structures in the universe. For instance, the presence of such self-interactions in the early universe can delay the onset of neutrino free-streaming, thereby altering the anisotropies in the CMB spectrum\,\cite{Follin:2015hya, Choi:2018gho, Baumann:2017lmt, Kreisch:2019yzn}. Furthermore, the growth of matter fluctuations is sensitive to these interactions through their effects on gravitational potentials\,\cite{Ma:1995ey, Dodelson:2003ft}. The absence of any indication of such new physics in cosmological observations places constraints on the mass and coupling of the mediator particle. Since all relevant cosmological observations pertain to the epoch with $T \leq 1\,\mathrm{MeV}$, whereas $m_{Z'} \geq 6.5\,\mathrm{MeV}$ as discussed above, the constraint is typically expressed as $\tilde{g}^2/m_{Z'}^2 \lesssim 10^{-4}\,\mathrm{MeV}^{-2}$\,\cite{Das:2020xke}. Meanwhile, neutrino self-interactions may also affect the neutrino energy distribution at the epoch of BBN, modifying both the expansion rate of the universe and the neutron–proton conversion processes. As a result, they can influence the effective number of relativistic degrees of freedom and the primordial abundances of light elements\,\cite{Blinov:2019gcj, Grohs:2020xxd, Huang:2021dba}. However, the impact of such interactions on these observables remains within the current uncertainties of cosmological measurements. In addition, non-standard neutrino interactions with WIMPs, which induce scattering between neutrinos and WIMPs, also affect the cosmological observables mentioned above. These interactions suppress primordial density fluctuations by erasing structures smaller than the collisional damping scale, thereby producing observable signatures in the CMB and the matter power spectrum, which leads to an upper bound on the parameters as $\tilde{g}^2 g_\varphi^2 / (m_{Z'}^4 m_\varphi) \lesssim 4.5 \times 10^{-8}\,\mathrm{MeV}^{-5}$\,\cite{Wilkinson:2014ksa, Escudero:2015yka, Bertoni:2014mva}.

\subsection{Constraints from astrophysical observations}
\label{subsec: astrophysical constraints}

The light neutrinophilic WIMP discussed in this article produces a monochromatic neutrino signal via its annihilation process $\varphi \varphi^* \to \nu \bar{\nu}$ in space, which can potentially be detected by a neutrino observatory. The flux of this signal is given by the following formula\,\cite{Arguelles:2019ouk}:
\begin{align}
    \frac{d \Phi_{\nu\,(\bar{\nu})}}{d E_{\nu\,(\bar{\nu})}} =
    \frac{1}{16 \pi m_\varphi^2}
    \sum_i \langle \sigma v \rangle
    {\rm Br}(\varphi \varphi^* \to f)
    \left.
    \frac{d N_{\nu\,(\bar{\nu})}}{d E_{\nu\,(\bar{\nu})}}
    \right|_f\,J(\Omega).
\end{align}
Here, $E_{\nu\,(\bar{\nu})}$ denotes the energy of the produced neutrinos (anti-neutrinos), ${\rm Br}(\varphi \varphi^* \to f)$ is the branching fraction into the final state $f$, and $\langle \sigma v \rangle$ is the velocity-averaged annihilation cross-section, computed using the velocity distribution $f(v; v_0) \simeq 4\pi^{-1/2} v^2 e^{-v^2/v_0^2} / v_0^3$, for which the average velocity is $\langle v \rangle = 2 v_0 / \sqrt{\pi}$. For a given final state $f$, the number of neutrinos (anti-neutrinos) produced per annihilation with energy $E_{\nu\,(\bar{\nu})}$ is given by
\begin{align}
    \frac{d N_{\nu\,(\bar{\nu})}}{d E_{\nu\,(\bar{\nu})}} =
    \delta(E - m_N),
\end{align}
when the WIMP annihilates directly into a pair of neutrinos, whereas a box-shaped spectrum is obtained when neutrinos are produced through WIMP annihilation into a pair of mediator particles followed by the mediator decays into neutrino pairs, $\varphi \varphi^{*} \to Z' Z' \to \nu \bar{\nu}\,\nu \bar{\nu}$.

Meanwhile, $J(\Omega)$ is called the $J$-factor, determined by integrating the square of the dark matter density, $\rho_{\rm DM}^2(\vec{r})$, along the line of sight in the direction $\Omega$, over a solid angle $\Delta \Omega$:
\begin{align}
    J(\Omega) \equiv
    \int_{\Delta \Omega} d\Omega \int_{l.o.s} dl\,\rho_{\rm DM}^2(\vec{r}).
    \label{eq: J-facgtor}
\end{align}
Among the various sources of neutrino signals, the contribution from our Milky Way galaxy is expected to be the strongest. As discussed in Section~\ref{subsec: self-scattering}, both smaller structures such as dwarf galaxies and larger ones like small clusters tend to exhibit cored dark matter profiles. However, recent observations have revealed significant variation in dark matter profiles across galaxies: while some show strongly cored profiles, others exhibit cuspy ones. In particular, galaxies with scales comparable to that of the Milky Way display a wide range of profiles. This phenomenon is known as the ``diversity problem'' of galaxies\,\cite{Oman:2015xda}. Within the traditional WIMP framework, this diversity may be attributed to the varying impact of baryonic effects on galaxy formation. In contrast, in the framework of self-interacting dark matter, the dark matter profile is significantly shaped by self-interactions when the baryon density at the galactic center is sufficiently high, leading to a wide range of dark matter distributions\,\cite{Kamada:2016euw}. Since our galaxy has a high baryon density at its center, the dark matter distribution is expected to be affected accordingly in either case. Therefore, we estimate the $J$-factor of the Milky Way based on observational data, such as measurements of the circular velocities of stars and gas. Although the dark matter profile at the galactic center is constrained by observations, it is not well determined and still has large uncertainties\,\cite{Benito:2019ngh, Benito:2020lgu}, which allows us to consider various profile shapes without conflicting with the data. To ensure the robustness of our analysis, we adopt the most conservative value of the $J$-factor that remains consistent with the data at the 95\,\% C.L.: $J(\Omega) = 1.2 \times 10^{23}$\,GeV$^2$\,cm$^{-5}$, which is obtained assuming a cored dark matter profile\,\cite{Burkert:1995yz, Benito:2020lgu}. Here, we calculate the $J$-factor over the entire sky, i.e., the $\Omega$-integration in Eq.\,(\ref{eq: J-facgtor}) is performed over the full solid angle.

The resulting monochromatic neutrino flux is compared with the upper limits reported by various neutrino observation experiments, thereby placing constraints on the annihilation cross-section of the neutrinophilic WIMP. The upper limit on the cross-section (times the relative velocity between incident WIMPs) is $10^{-22}$--$10^{-25}\,{\rm cm}^3/{\rm s}$ when the mass of the WIMP is $1$--$100\,{\rm MeV}$\,\cite{Arguelles:2019ouk}. This limit is above the canonical cross-section of $\sim 10^{-26}\,{\rm cm}^3/{\rm s}$ required to satisfy the relic abundance condition, and does not impose a severe constraint on the neutrinophilic WIMP model, as long as the cross-section does not change significantly in the non-relativistic regime. On the other hand, if the cross-section exhibits a sizable velocity dependence in the non-relativistic region, such as in the case of an $s$-channel resonance relevant to the self-scattering condition shown in Eq.\,(\ref{eq: Thermally Averaged Ann Xsection}), the neutrino observation limits impose meaningful constraints on the WIMP model, as discussed in the next section.

On the other hand, neutrino self-interactions mediated by a neutrinophilic gauge boson can also be constrained by astrophysical observations. One notable bound arises from the detection of neutrinos from the supernova SN1987A: in the core of an SN, sufficiently strong self-interactions can trigger processes such as $2\nu \to 4\nu$, thereby reducing the energy transferred by neutrinos to the stalled shock wave and potentially hindering its revival. This consideration leads to the constraint $\tilde{g} \lesssim 0.026$ for $m_{Z'} < 10\,\mathrm{MeV}$ and $\tilde{g} \lesssim 0.06 \times (m_{Z'}/100\,\mathrm{MeV})$ for $m_{Z'} > 100\,\mathrm{MeV}$\,\cite{Shalgar:2019rqe}. Another constraint arises from the detection of high-energy extragalactic neutrinos associated with identified astrophysical sources. If the source luminosity can be independently inferred, the intrinsic neutrino luminosity can also be estimated. In such cases, neutrino self-interactions that lead to a mean free path, due to scattering with the cosmic neutrino background, shorter than the distance to the source are excluded, provided that the observed flux is consistent with expectations. Using the blazars TXS\,0506+056 and PKS\,0735+178, the active galaxy NGC~1068, and the KM3-230213A neutrino event, the resulting constraint on the coupling $\tilde{g}$ is generally weaker than, or at most comparable to, that obtained from SN1987A addressed above\,\cite{Kelly:2018tyg, Bustamante:2020mep, Petropavlova:2024wia}. Moreover, further constraints can be obtained by observing the diffuse neutrino flux from a population of sources. While astrophysical acceleration mechanisms typically predict a power-law spectrum, self-interactions would introduce spectral features such as dips and bumps, as the scattering is resonantly enhanced at particular energies. Based on six years of publicly available IceCube HESE data, upper limits on the coupling $\tilde{g}$ have been obtained in the mass range $m_{Z'} = 1$--$100\,\mathrm{MeV}$, surpassing the SN1987A bound for $m_{Z'} = 1$--$8\,\mathrm{MeV}$, where $\tilde{g} < 0.016$--$0.004$\,\cite{Bustamante:2020mep}.

Additionally, high-energy neutrinos observed at IceCube from active galactic nuclei (such as TXS~0506+056 and NGC~1068) place upper limits on the interaction between neutrinos and WIMPs, since the observed neutrinos propagate through the dark matter halos hosting these galactic nuclei. These limits are expressed in terms of the neutrino--WIMP scattering cross-section as a function of the WIMP mass and the integrated dark matter column density along the line of sight to IceCube\,\cite{Cline:2023tkp, Fujiwara:2023lsv, Heston:2024ljf}. The constraint, however, depends sensitively on the dark matter halo profile of the host galaxy; for example, a spike profile leads to a stronger limit on the cross-section. Given that the scattering cross-section is suppressed by the square of the incoming neutrino energy, the limit derived from these observations becomes less stringent than those from the cosmological observations discussed in the previous subsection, especially if a less aggressive spike profile is assumed for the host galaxy.

\subsection{Constraints from terrestrial experiments}

The neutrinophilic WIMP and the mediator particle $Z'$ are also potentially constrained by several terrestrial experiments. One such example is direct dark matter detection at underground laboratories. As discussed in Section\,\ref{sec: scenario}, particularly around Eq.\,(\ref{eq: scattering cross-section}), the neutrinophilic WIMP can scatter off nuclei via its interaction with neutrons. Meanwhile, current results from direct detection experiments place an upper limit on the dark matter--nucleon scattering cross-section, under the assumption that dark matter interacts equally with protons and neutrons. The most stringent constraint arises from the XENON1T experiment, which utilizes the Migdal effect, in the mass range of 0.1--1\,GeV. This experiment probes the scattering between dark matter and the natural xenon nucleus, which has an atomic number of 54 and a standard atomic weight of $131.3$. This yields the following limit:
\begin{align}
    \sigma (\phi^{(*)} n \to \phi^{(*)} n) \lesssim
    \left(\frac{131.3}{131.3 - 54 \times 1.007}\right)^2
    \sigma ({\rm DM\,N} \to {\rm DM\,N})|_{\rm Limit},
\end{align}
where the number 1.007 denotes the atomic weight of a proton, and $\sigma ({\rm DM\,N} \to {\rm DM\,N})|_{\rm Limit}$ represents the upper limit on the scattering cross-section between dark matter (DM) and a nucleon (N), as obtained by the XENON1T experiment. Since $\sigma ({\rm DM\,N} \to {\rm DM\,N})|_{\rm Limit} \lesssim 10^{-39}\,(1\,{\rm GeV}/m_{\rm DM})^5\,{\rm cm}^2$ for a dark matter mass of $m_{\rm DM} = 0.1$--$1\,{\rm GeV}$\,\cite{XENON:2019zpr}, this constraint can be translated into an upper limit on the model parameters of a neutrinophilic WIMP as $\tilde{g}^2\,g_\varphi^2\,(m_\varphi/1\,{\rm GeV})^7\,(m_{Z'}/1\,{\rm GeV})^{-4} \lesssim 8 \times 10^{-12}$, based on the scattering cross-section\,(\ref{eq: scattering cross-section}).

%About CEνNS
%gBL < 2x10^-5 when mZ' < O(10)MeV (20MeV)
%gBL < 10^-4 (mZ'/100MeV) when O(10)MeV mZ'>20MeV
%Ge:Z=32,N=38,40,41,42,44@20.57%,27.45%,7.75%,36.50%, 7.73%
%Average N=40.705 [Molar mass=72.64->72.64g/mol] -> 0.56
%Cs:Z=55,N=78 & I:Z=53,N=74 -> 0.58
%Ar:Z=18,N=22 -> 0.55
Another example is coherent elastic neutrino--nucleus scattering (CE$\nu$NS). In addition to the SM process mediated by $Z$-boson exchange, neutrinos can scatter off nuclei via $Z'$, which couples to both neutrinos and neutrons. Among CE$\nu$NS experiments, the most stringent bounds on such a neutrinophilic mediator currently come from COHERENT\,\cite{COHERENT:2017ipa, COHERENT:2020iec, COHERENT:2024axu} for $m_{Z'} > \mathcal{O}(10)\,\mathrm{MeV}$, while CONUS+\,\cite{Ackermann:2025obx} provides the strongest constraints in the light mediator regime, $m_{Z'} < \mathcal{O}(10)\,\mathrm{MeV}$. The resulting limits on the coupling are $\tilde{g} \lesssim 3.6\times10^{-5}$ for $m_{Z'} < \mathcal{O}(10)\,\mathrm{MeV}$ and $\tilde{g} \lesssim 1.8\times10^{-4}\,(m_{Z'}/100\,\mathrm{MeV})$ for $m_{Z'} > \mathcal{O}(10)\,\mathrm{MeV}$\,\cite{Chattaraj:2025fvx, AtzoriCorona:2025ygn, TEXONO:2025sub}. Note that, as in the case of direct dark matter detection discussed above, coherent nuclear scattering via $Z'$ proceeds through its coupling to neutrons. Consequently, the effective coupling to target nuclei such as Ge and CsI is suppressed by a factor of $N/(Z+N) \sim 0.56$ relative to a mediator that couples universally to all quarks (i.e., a pure $U(1)_{\rm B-L}$ gauge boson), where $Z$ and $N$ denote the proton and neutron numbers of the target nuclei, respectively.

%Double beta decay (with a mediator emission): superweak. \\
%$[G_S < 3.2 \times 10^8 G_F$ w/ $G_S = g_nu^2/(t + m_{Z'}^2) \to g_\nu < \mathcal{O}(1)$ for $m_{Z'} < 100$\,MeV.] \\
%Charged meson decays (radiative emission): $\pi$ \& K stronger than D \& B. \\
%$[$Longitudinal enhancement.$]$ \\
%Pi decay: $\pi^0 \to \gamma Z'$ \cite{cortina2019search}
%Z decay ($Z \to \nu \bar{\nu} Z'$): $g_\nu < 0.54 @ m_{Z'} < 10$\,GeV. \\
%W decay ($W \to \ell \nu Z'$): Expected to be comparable to $Z$ (no analysis). \\
%Tau decay ($\tau \to \ell \nu \nu Z'$): weak. \\
%Tritium decay($^3$H $\to$ $^3$He $e \bar{\nu}_e Z'$): Only for tiny $m_{Z'}$. \\
%Virtual emission of $\pi$, K, Z, W: superweak. \\
%Neutrino decay ($\nu_j \to \nu_i + Z'$): Only for tiny $m_{Z'}$. \\ 
%Neutrino bound state: Only for tiny $m_{Z'}$. \\
%Dune @ Future (mono-neutrino): $\tilde{g} < 10^{-2}$--$5 \times 10^{-3}$ when $m_{Z'} < {\cal O}(100)$\,MeV. \\ 
%LHC FPF (mono-neutrino): $\tilde{g} < 0.1$ when $m_{Z'} > {\cal O}(1)$\,GeV. \\
%W fusion process $pp \to WWX -> \to ee\nu\nu X$: $\tilde{g} \lesssim O(1)$ for up to ${\cal O}(100)$\,GeV mass. \\
%Higgs decay ($h \to Z Z'$, etc.): $\tilde{g} \lesssim 0.3$. Model-dependent? \\
%UV particle productions. \\
Additionally, the neutrinophilic mediator $Z'$ is subject to a wide range of laboratory constraints, including searches for double-beta decay, rare decays of SM particles (mesons, charged leptons, and the $W$, $Z$, and Higgs bosons), as well as neutrino-scattering and collider experiments. Among these, the most stringent bounds currently arise from rare charged-meson decays, particularly the process $K^\pm \to \pi^\pm + Z'$\,\cite{Beacham:2019nyx, Goudzovski:2022vbt}. Consequently, the coupling is constrained to $\tilde{g} \lesssim 2 \times 10^{-2}\,(1\,\mathrm{MeV}/m_{Z'})$, except in the mass range $m_{Z'} = 110$--$150$\,MeV\,\cite{NA62:2025upx}, where the background from the hadronic decay $K^\pm \to \pi^\pm \pi^0$ dominates, requiring a different analysis strategy and yielding a weaker bound of $\tilde{g} \lesssim 10^{-3}$\,\cite{NA62:2019meo, NA62:2020pwi}.
%Among these, the most stringent bounds on the mediator $Z'$ currently arise from rare decays of charged mesons ($\pi^\pm$ and $K^\pm$), particularly from processes in which a $Z'$ is radiated off the final-state (anti)neutrino, i.e., $\pi^\pm,\, K^\pm \to \ell^\pm + \nu\,(\bar{\nu}) + Z'$, where $\ell$ denotes an SM charged lepton ($e$ or $\mu$)\,\cite{xxx}. The branching fractions of these decay modes are enhanced by the longitudinal polarization of a light $Z'$ (Goldstone-equivalence effect), scaling as $m_{K,\,\pi}^2/m_{Z'}^2$ for $m_{Z'} \ll m_{K,\,\pi}$, modulo phase-space and form-factor effects. Consequently, one finds a coupling constraint $\tilde{g} \lesssim 10^{-5}\,(m_{Z'}/1\,\mathrm{MeV})$ for $m_{Z'} \lesssim 300\,\mathrm{MeV}$\,\cite{xxx}. These limits are largely model-independent when $Z'$ couples predominantly to neutrinos.

%%%%%%%%%%%%%%%%%%%%%%%%%%%%%%%%%%%%%%%%
%%%%%%%%%%% Variable regions %%%%%%%%%%%
%%%%%%%%%%%%%%%%%%%%%%%%%%%%%%%%%%%%%%%%
\section{Quantitative analysis of the neutrinophilic WIMP}
\label{sec: results}

\begin{table}[h]
    \centering
    \begin{tabular}{r|llccc}
        & $m_\varphi$ & $m_{Z'}$ & $\tilde{g}$ & $g_\varphi$ & $\lambda_\varphi$ \\
        \hline
        Parameter set 1 & 40\,MeV & 85\,MeV & $9 \times 10^{-6}$ & 1.8 & --- \\
        \ditto \qquad \quad 2 & 40 \quad \ditto & 10 \quad \ditto & --- & $5.2 \times 10^{-3}$ & --- \\
        \hline
    \end{tabular}
    \caption{\small\sl Representative typical model parameter sets that satisfy the relic-abundance condition as well as relevant cosmological, astrophysical, and terrestrial constraints. See the text for more details.}
    \label{tab: viable sets}
\end{table}

%mphi>5.3MeV or mZ'>6.5MeV.
%mphi<139MeV.
%gtilde^2/mZ'^2<10^-4MeV^-2.
%gtilde^2gphi^2/(mZ'^4mphi)<4.5x10^-8MeV^-5.
%ID using neutrino observations.
%gtilde<0.026@mZ'<10MeV & <0.06(mZ'/100MeV)@mZ'>100MeV.
%gtilde<0.016-0.004@mZ'=1-8MeV
%gtilde^2gphi^2(mphi/1GeV)^7(mZ'/1GeV)^-4<8x10^-12
%gtilde<3.6x10^-5@mZ<20MeV & <1.8x10^-5(mZ'/10MeV)@mZ'>20MeV
We examine the status of the neutrinophilic scalar WIMP scenario by presenting viable parameter sets that satisfy the relic-abundance requirement for the present dark matter density, while remaining consistent with cosmological, astrophysical, and terrestrial constraints on both the dark matter and the mediator particles. Representative parameter sets are listed in Table~\ref{tab: viable sets}. In parameter set~1, the WIMP annihilates into a pair of neutrinos via $s$-channel exchange of the mediator particle $Z'$, whereas in parameter set~2, the WIMP instead annihilates into a pair of mediator particles. Since, in the latter case, the annihilation cross-section does not depend on the coupling $\tilde{g}$, we do not specify its value in the table; it can be chosen small enough to evade the constraints discussed in the previous section, while remaining large enough to maintain thermal equilibrium with the SM particles at freeze-out\,\cite{Watanabe:2025pvc}. Moreover, because the WIMP self-interaction coupling $\lambda_\varphi$ is irrelevant for both the relic-abundance requirement and those constraints, we omit its value for both parameter sets in the table. In fact, a wide region of parameter space remains viable thanks to the neutrinophilic nature of the WIMP in both scenarios, namely, annihilation into a neutrino pair or into a mediator pair. How such an extensive parameter space can be probed in future experiments remains an important open question, which we leave for future investigation.

On the other hand, while the neutrinophilic scalar WIMP model is interesting in its own right, we also examine the parameter region where the self-scattering condition discussed in Sec.\,\ref{subsec: self-scattering} is also imposed in addition to the previously considered conditions and constraints. This enables us to address the small-scale structure problem of the universe and to establish a more well-motivated scenario. To identify such a region, we adopt the following analysis. First, we impose the requirement that the chosen set of model parameters yields a cross-section consistent with the self-scattering data in Fig.\,\ref{fig: self-scattering}, using the likelihood function,
\begin{eqnarray}
    -2\ln {\mathcal L}_{\rm SS} =
    \sum_i
    \left\{
        \frac{(\text{Center value})_i
        -\log_{10}[ \langle \sigma_{\rm SS} v \rangle_{v_{0,i}}/m_\varphi]}
        {(\text{Error bar})_i}
    \right\}^2,
\end{eqnarray}
where the index $i$ labels a galaxy or a cluster in the data, each characterized by a central value and an associated error in the figure, and $v$ denotes the relative velocity between incoming WIMPs. A set of model parameters $(m_\varphi,\,v_{\rm R},\,g_\varphi,\,\sigma_0)$ is considered viable if it is not excluded by the data at a confidence level exceeding $95\%$.\footnote{
    The data should be interpreted with caution, as additional systematic uncertainties may arise in the extraction of cross-sections from kinematical observations\,\cite{Sokolenko:2018noz}. Nevertheless, we employ these data to delineate the region of parameter space in which the WIMP scenario is favored as a resolution of the diversity problem.}
Next, we employ the following likelihood function to impose the relic abundance condition, ensuring that the chosen set of model parameters remains consistent with the present-day dark matter density:
\begin{eqnarray}
    -2\ln {\mathcal L}_{\rm RA} =
    \left(
        \frac{\Omega_{\rm DM}\,h^2 - \Omega_{\rm TH}\,h^2}{0.15\,\Omega_{\rm DM}\,h^2}
    \right)^2,
\end{eqnarray}
where $\Omega_{\rm TH} h^2$ denotes the WIMP relic abundance predicted by the neutrinophilic WIMP model. We compute this abundance with the DRAKE code\,\cite{Binder:2021bmg}, which accounts for the early kinetic decoupling effect. During freeze-out, the WIMP is expected to remain in self-thermal equilibrium owing to the self-interactions discussed above; consequently, its momentum distribution is well approximated by a Maxwell-Boltzmann form. In our analysis, we assign a theoretical uncertainty of $15\%$ to $\Omega_{\rm TH} h^2$, obtained from three sources added in quadrature. First, using the latest relativistic degrees of freedom in place of the code default induces an uncertainty of ${\cal O}(0.1)\%$\,\cite{Saikawa:2020swg}. Second, the scattering collision term in the Boltzmann equation is approximated by a Fokker--Planck operator in the code\,\cite{Rosenbluth:1957zz, Bringmann:2006mu}, leading to an uncertainty of about $10\%$. Third, freeze-out occurs near neutrino decoupling, for which a fully consistent treatment would require solving three coupled Boltzmann equations (for the WIMP, the electromagnetic $e^\pm/\gamma$ plasma, and neutrinos). Since this is not included in the code, we estimate an additional uncertainty of up to $10\%$, as the WIMP interacts exclusively with neutrinos\,\cite{Li:2023puz}. Then, a parameter point is deemed viable if it is not excluded at more than $95\%$ confidence level by the observational result $\Omega_{\rm DM} h^2 = 0.120 \pm 0.001$\,\cite{Garcia-Garcia:2024gzy}.
%This approximation is valid when the typical momentum transfer per scattering is smaller than the average momentum of the WIMP in thermal equilibrium. This situation occurs, for example, when the WIMP is non-relativistic and scatters with a lighter particle (such as neutrinos in our scenario) in the thermal bath.
%Based on this, the scattering cross-section between the WIMP and neutrinos is assumed to be negligibly small during the freeze-out epoch to reduce the computational cost of evaluating $\Omega_{\rm TH}\,h^2$. This assumption is justified because we focus on the resonance region, where the coupling $\tilde{g}$ is highly suppressed, as discussed above. Indeed, we numerically verified with several sets of model parameters that the relic abundance computed by including the scattering process is consistent with that computed without it; the two results agree within a few percent.

\begin{figure}[t]
    \centering
    \includegraphics[width=0.95\linewidth]{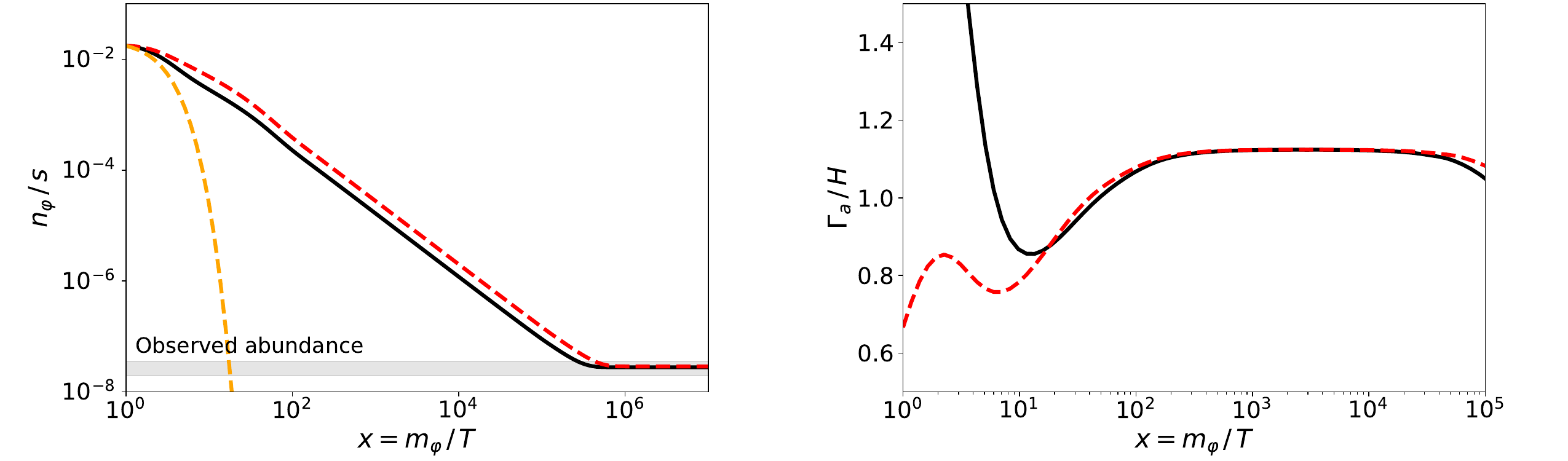}
    \caption{\small \sl 
    {\bf Left panel:} The number density of the WIMP ($n_\varphi$), normalized by the entropy density ($s$), is shown as a function of the universe's temperature ($T$). The prediction using the parameter set $(m_\varphi, v_R, g_\varphi, \tilde{g}) = (7.7\,\text{MeV}, 108\,\text{km/s}, 2.4 \times 10^{-3}, 9.1 \times 10^{-11})$ is represented by a black solid line, while the one using $(7.7\,\text{MeV}, 65\,\text{km/s}, 3.1 \times 10^{-3}, 5.2 \times 10^{-11})$ is represented by a red dashed line. Both lines assume the initial condition $f_\varphi(p) = f_{\rm eq}(p)$ (i.e., $n_\varphi = n_{\rm eq}$) at $T = m_\varphi$ and satisfy the relic abundance condition, which is represented by the light-gray band. The comparison between the WIMP annihilation rate ($\Gamma_a$) and the expansion rate of the universe ($H$) as a function of $T$ is shown using the same parameter sets as in the left panel. See the main text for more details.}
    \label{fig: relic abundance}
\end{figure}

Here, an interesting feature of the temperature evolution of $n_\varphi/s$ (used to compute $\Omega_{\rm TH} h^2$) is that, unlike in the traditional WIMP scenario, the freeze-out process is not instantaneous due to the presence of an $s$-channel resonance, where $n_\varphi$ and $s$ denote the WIMP and the entropy densities, respectively. In this case, the WIMP annihilation cross-section increases as the temperature of the universe decreases. As a result, the abundance continues to decrease even after the system departs from equilibrium, as shown in the left panel of Fig.\,\ref{fig: relic abundance}. Consequently, for certain choices of model parameters, the WIMP annihilation rate $\Gamma_a \sim \langle \sigma v \rangle n_\varphi$ can remain comparable to or even smaller than the expansion rate of the universe (i.e., the Hubble parameter $H \propto x^{-2}$) during the epoch $x \gtrsim 1$, as illustrated by the red dashed line in the right panel of the figure. Meanwhile, according to Eq.\,(\ref{eq: Thermally Averaged Ann Xsection}), the thermally averaged cross-section in the early universe ($x \lesssim 1$) can be approximated as $\langle \sigma v \rangle \simeq 3 \pi \tilde{g}^2 x^4 /(4 m_\varphi^2)$. Assuming that the WIMP density follows the equilibrium distribution at $x \lesssim 1$ (i.e., $n_{\rm DM} \propto x^{-3}$), the annihilation rate scales as $\Gamma_a \propto x$. This implies that the ratio of the annihilation rate to the expansion rate behaves as $\Gamma_a/H \propto x^3$ during this epoch. Therefore, if $\Gamma_a/H \lesssim 1$ at $x \sim 1$, the annihilation rate may remain smaller than the expansion rate throughout the entire thermal history of the universe, thereby losing the attractive feature of the thermal freeze-out scenario.\footnote{
    The attractive feature refers to the fact that the thermal relic abundance is determined independently of the initial condition of the WIMP density in the very early universe, e.g., its value immediately after inflation.} 
In our analysis, we compute the relic abundance $\Omega_{\rm TH} h^2$ under the assumption that the WIMP is in equilibrium with the SM particles at $T = m_\varphi$, i.e., $f_\varphi(p, T)\big|_{T = m_\varphi} = f_{\rm eq}(p, T) \equiv \exp[-(p^2 + m_\varphi^2)^{1/2}/T]$, where $f_\varphi$ is the occupation number of the WIMP. Then, we further impose the condition $\Gamma_a > H$ at $x = 1$.

On the other hand, the coupling $\tilde{g}$ must be highly suppressed to satisfy both the self-scattering and relic-abundance requirements. For example, $\tilde{g}$ must be of order $10^{-10}$ when $m_\phi \sim 10\,\text{MeV}$, as shown in Fig.~\ref{fig: relic abundance}. As a result, most of the constraints discussed in the previous section, cosmological and astrophysical bounds on neutrino--neutrino and neutrino--WIMP interactions, as well as limits from terrestrial experiments, are automatically fulfilled and can be neglected in determining the viable parameter space. The relevant constraints are therefore those from CMB observations on the effective number of relativistic degrees of freedom $N_{\rm eff}$ and on the WIMP annihilation cross-section into electrically interacting particles (section~\ref{subsec: cosmological constraints}), together with those from indirect dark-matter searches at neutrino observatories (section~\ref{subsec: astrophysical constraints}). The former restricts the WIMP mass to $5.8\,\mathrm{MeV} \leq m_\varphi \leq 139\,\mathrm{MeV}$, while the latter constrains the present-day WIMP annihilation cross-section into neutrinos. We thus scan the parameter space over this mass interval, imposing the self-scattering and relic-abundance conditions, and identify the consistent region. Finally, we compare this region with the indirect-detection bounds to delineate the viable parameter space.

\begin{figure}[t]
    \centering
    \includegraphics[width=0.8\linewidth]{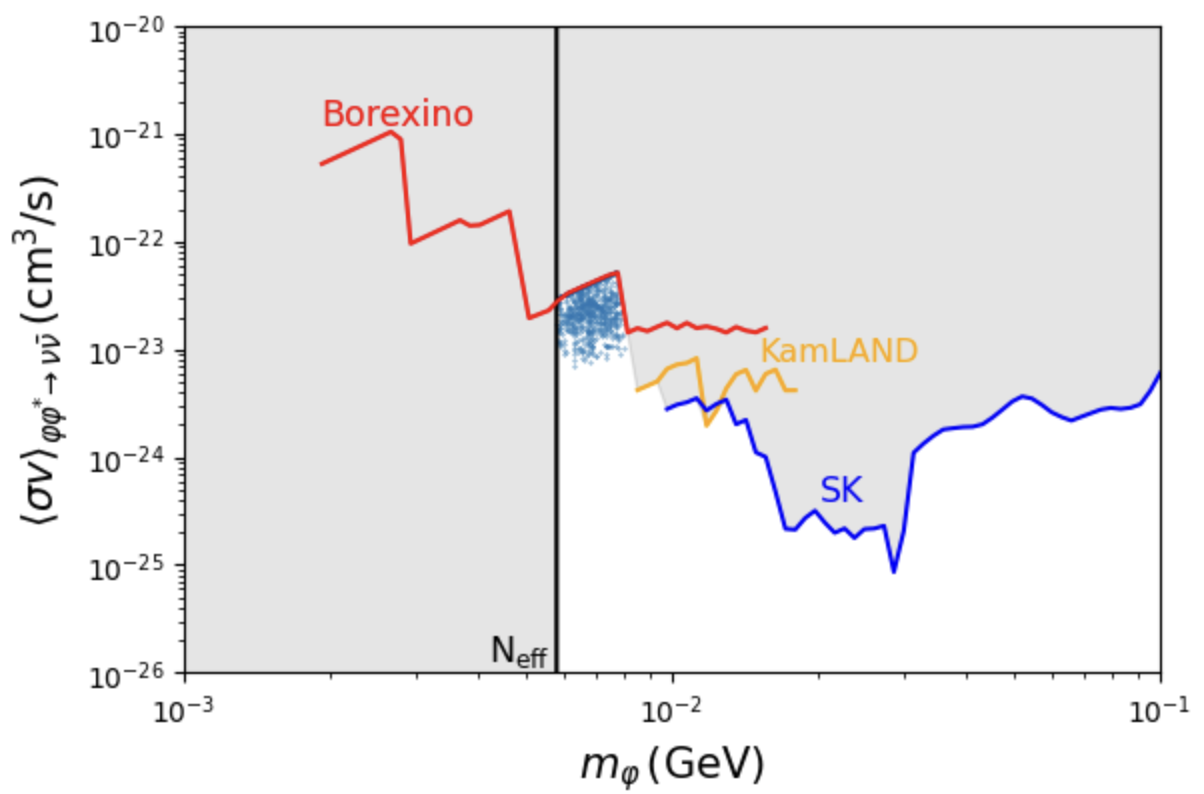}
    \caption{\small \sl Regions of parameter space that satisfy the relic-abundance and self-scattering conditions, while being consistent with existing cosmological, astrophysical, and terrestrial constraints, shown in the $(m_\varphi,\langle\sigma v(\varphi\varphi^{*}\!\to\!\nu\bar{\nu})\rangle_{0})$ plane. We take the WIMP velocity dispersion to be $v_{0}\simeq 400\,\mathrm{km/s}$. Shaded regions indicate exclusions from indirect searches at neutrino observatories, together with the CMB bound from $N_{\rm eff}$ (implying $m_\varphi \ge 5.8\,\mathrm{MeV}$). See the main text for more details.}
    \label{fig: results}
\end{figure}

The results of this analysis are presented in Fig.\,\ref{fig: results}, which shows the regions of parameter space satisfying both conditions, projected onto the plane spanned by $m_\varphi$ and the present-day annihilation cross-section into a neutrino pair, $\langle \sigma v(\varphi\varphi^{*}\to \nu\bar{\nu})\rangle_{0}$. Since the neutrino flux from WIMP annihilation in the Galactic center dominates the signal, we take the WIMP velocity dispersion to be $v_{0}\simeq 400\,\mathrm{km/s}$, as suggested by recent N-body simulations\,\cite{Lacroix:2020lhn, McKeown:2021sob}. Constraints from indirect searches at neutrino observatories, together with the CMB bound from $N_{\rm eff}$ (i.e., $m_\varphi \ge 5.8\,\mathrm{MeV}$), are indicated by the shaded regions, and one finds that Borexino\,\cite{Borexino:2019wln}, KamLAND\,\cite{KamLAND:2011bnd}, and Super-Kamiokande\,\cite{Wan:2018ndu} provide the most stringent bounds in the $\mathcal{O}(10)\,\mathrm{MeV}$ range. It is evident from the figure that, because the self-scattering condition is realized through an $s$-channel resonance, the predicted annihilation cross-section is much larger than the so-called ``thermal'' cross-section, which refers to the value required to reproduce the observed dark matter abundance when $\langle\sigma v\rangle$ is approximately velocity independent, in contrast to the strongly velocity-dependent case considered here. Consequently, the WIMP mass is constrained to be below $\sim 8\,\mathrm{MeV}$, and combining this with the CMB lower limit we find that a viable region satisfying both the relic-abundance and self-scattering conditions still exists, albeit in the narrow window $5.8\,\mathrm{MeV}\le m_\varphi \le 8\,\mathrm{MeV}$. Interestingly, much of this well-motivated region may be probed with existing KamLAND data, which have not yet been analyzed for this class of dark-matter searches.\footnote{This statement is based on private communication with a colleague at the authors' institution.}

\section{Summary}
\label{sec: summary}

Thermal dark matter remains a compelling explanation of the relic abundance, but light WIMPs (sub-GeV dark matter) are strongly constrained because CMB anisotropies limit energy injection from annihilation at recombination, which disfavors the standard $s$-wave freeze-out scenario into electromagnetically interacting final states. A natural way to evade this bound is to consider neutrinophilic WIMPs, although realizing dominant couplings to neutrinos without inducing comparable couplings to charged leptons is nontrivial. Since light WIMPs are typically SM singlets connected to the SM through an additional mediator, we proposed a new neutrinophilic WIMP scenario based on an extra Abelian gauge symmetry $\mathrm{U}(1)_{\mathrm{B}-\mathrm{L}+x\mathrm{Y}}$: for $x\simeq -1/c_W^2$, the corresponding gauge boson $Z'$, the mediator particle, couples predominantly to neutrinos, while its couplings to both chiralities of charged leptons are strongly suppressed below the electroweak scale. In the sub-GeV regime, hadronic channels are kinematically closed, so the WIMP annihilates essentially only into neutrinos; after systematically incorporating constraints from cosmological limits on $\Delta N_{\rm eff}$, cosmological and astrophysical bounds on neutrino self-interactions and neutrino--WIMP scattering, and terrestrial probes including direct detection, CE$\nu$NS measurements, and accelerator searches, we found a broad and robustly viable region that reproduces the observed relic abundance via conventional freeze-out, with representative benchmark sets given in Table~\ref{tab: viable sets}. We then examined the more predictive region where the model also yields self-interacting dark matter, relevant for alleviating the small-scale structure problem, and performed likelihood-based fits to self-scattering data, together with the relic density computed with \texttt{DRAKE} (including early kinetic decoupling and associated theory uncertainties). Combining all requirements, we identified a viable window $5.8~\mathrm{MeV}\le m_\varphi\le 8.3~\mathrm{MeV}$ that satisfies both the relic-abundance and self-scattering conditions while remaining consistent with current indirect limits, with Borexino, KamLAND, and Super-Kamiokande being the most relevant. Much of this narrow but well-motivated region may be testable with existing KamLAND data, motivating a dedicated reanalysis and complementary future searches.

\appendix

%%%%%%%%%%%%%%%%%%%%%%%%%%%%%%%%%%%%%
%%%%%%%%%%% B_munu V^munu %%%%%%%%%%%
\section{Kinematic mixing term}
\label{app: kinetic mixing}

We confirm that the $\mathrm{U(1)_{B-L+xY}}$ extension of the SM, which includes the kinetic mixing term between the field strength tensors of the $\mathrm{U(1)_{B-L+xY}}$ and hypercharge gauge bosons, namely the term ${\cal L} \supset -(\xi/2)\,B_{\mu\nu}V^{\mu\nu}$, becomes phenomenologically indistinguishable from the $\mathrm{U(1)_{B-L+xY}}$ extension without such a kinetic mixing term when both the gauge coupling of $\mathrm{U(1)_{B-L+xY}}$, i.e., $\tilde{g}$, and the kinetic mixing parameter $\xi$ are sufficiently small\,\cite{Lee:2016ief, Fayet:2016nyc}.

In the U(1)$_{\rm B-L+xY}$ model, the covariant derivatives acting on the SM fermions, the SM Higgs boson, the right-handed neutrinos, and the new Higgs boson $S$ are generally given by
\begin{align}
    D_\mu =
    \partial_\mu
    - i
    \left(
        g_S G_\mu
        + g W_\mu
        + g' Y B_\mu
        + \tilde{g} q V_\mu
    \right),
\end{align}
where $G_\mu \equiv G_\mu^a T_G^a$ and $W_\mu \equiv W_\mu^a T_W^a$. By performing the redefinition of the gauge field $B_\mu$,
\begin{align}
    B_{\mu} \rightarrow
    B_{\mu} + \frac{\tilde{g}}{g'}yV_\mu,
\end{align}
with $y$ being a real parameter, the covariant derivative given in the above equation can be rewritten in a form corresponding to the U(1)$_{\rm B-L+(x+y)Y}$ extension of the SM as follows:
\begin{align}
    D_\mu \rightarrow D_\mu =
    \partial_\mu
    - i
    \left[
        g_S G_\mu
        + g W_\mu
        + g' Y B_\mu
        + g_{\rm B-L} (q + y Y) V_\mu
    \right].
\end{align}
This field redefinition changes the kinetic terms of the relevant gauge fields as follows:
\begin{align}
    \mathcal{L}_{\rm B-L} &\supset
    -\frac{1}{4}V_{\mu\nu} V^{\mu\nu}
    -\frac{1}{4}B_{\mu\nu} B^{\mu\nu}
    -\frac{\xi}{2} B_{\mu\nu} V^{\mu\nu}
    \nonumber \\
    &\rightarrow
    -\frac{1}{4}
    \left(
        1 
        + 2 \frac{\tilde{g}}{g'} y \xi
        + \frac{\tilde{g}^2}{g'^2} y^2
    \right) V_{\mu\nu} V^{\mu\nu}
    -
    \frac{1}{2}
    \left(
        \frac{\tilde{g}}{g'} y
        +\xi
    \right) B_{\mu\nu} V^{\mu\nu}
    -\frac{1}{4}B_{\mu\nu}B^{\mu\nu}.
\end{align}
Thus, choosing $y = - (g'/g_{\rm B-L})\,\xi$ eliminates the kinetic mixing, while the kinetic terms of the gauge fields remain canonical up to ${\cal O}(\xi)$, since $y \tilde{g} = {\cal O}(\xi)$. This implies that the U(1)$_{\rm B-L + xY}$ model with a kinetic mixing term in the Lagrangian is phenomenologically equivalent to the U(1)$_{\rm B-L + (x + y)Y}$ model without kinetic mixing, as long as $y \tilde{g}$ (i.e., $\xi$) is sufficiently small.

%%%%%%%%%%%%%%%%%%%%%%%%%%%%%%%%%%%%%%%
%%%%%%%%%%% Acknowledgments %%%%%%%%%%%
%%%%%%%%%%%%%%%%%%%%%%%%%%%%%%%%%%%%%%%
\section*{Acknowledgments}

T.~A. was supported by the Forefront Physics and Mathematics Program to Drive Transformation (FoPM), a World-leading Innovative Graduate Study (WINGS) Program at the University of Tokyo. S.~M. was supported by Grants-in-Aid for Scientific Research from the Ministry of Education, Culture, Sports, Science and Technology (MEXT), Japan, under Grant Nos.\ 24H00244 and 24H02244. Yu W. was supported by the JSPS Overseas Research Fellowships and the U.S.\ Department of Energy under Award No.\ DE-SC0009937. T.~A., S.~M., and Yuki W. were also supported by the World Premier International Research Center Initiative (WPI), MEXT, Japan (Kavli IPMU).

%%%%%%%%%%%%%%%%%%%%%%%%%%%%%%%%%%%
%%%%%%%%%%% Biblography %%%%%%%%%%%
%%%%%%%%%%%%%%%%%%%%%%%%%%%%%%%%%%%
\bibliographystyle{unsrt}
\bibliography{bibliography}

@article{Lee:2016ief,
    author = "Lee, Hye-Sung and Yun, Seokhoon",
    title = "{Mini force: The $(B-L)+xY$ gauge interaction with a light mediator}",
    eprint = "1604.01213",
    archivePrefix = "arXiv",
    primaryClass = "hep-ph",
    reportNumber = "CTPU-16-11",
    doi = "10.1103/PhysRevD.93.115028",
    journal = "Phys. Rev. D",
    volume = "93",
    number = "11",
    pages = "115028",
    year = "2016"
}

@article{Hisano:2017jmz,
    author = "Hisano, Junji",
    editor = "Davidson, Sacha and Gambino, Paolo and Laine, Mikko and Neubert, Matthias and Salomon, Christophe",
    title = "{Effective theory approach to direct detection of dark matter}",
    eprint = "1712.02947",
    archivePrefix = "arXiv",
    primaryClass = "hep-ph",
    doi = "10.1093/oso/9780198855743.003.0011",
    month = "12",
    year = "2017"
}

@article{Cirelli:2013ufw,
    author = "Cirelli, Marco and Del Nobile, Eugenio and Panci, Paolo",
    title = "{Tools for model-independent bounds in direct dark matter searches}",
    eprint = "1307.5955",
    archivePrefix = "arXiv",
    primaryClass = "hep-ph",
    reportNumber = "CP3-ORIGINS-2013-014, DIAS-2013-14, SACLAY-T13-022",
    doi = "10.1088/1475-7516/2013/10/019",
    journal = "JCAP",
    volume = "10",
    pages = "019",
    year = "2013"
}

@article{McDermott:2017qcg,
    author = "McDermott, Samuel D. and Patel, Hiren H. and Ramani, Harikrishnan",
    title = "{Dark Photon Decay Beyond The Euler-Heisenberg Limit}",
    eprint = "1705.00619",
    archivePrefix = "arXiv",
    primaryClass = "hep-ph",
    reportNumber = "ACFI-T17-08, YITP-SB-17-14",
    doi = "10.1103/PhysRevD.97.073005",
    journal = "Phys. Rev. D",
    volume = "97",
    number = "7",
    pages = "073005",
    year = "2018"
}

@article{Spergel:1999mh,
    author = "Spergel, David N. and Steinhardt, Paul J.",
    title = "{Observational evidence for selfinteracting cold dark matter}",
    eprint = "astro-ph/9909386",
    archivePrefix = "arXiv",
    doi = "10.1103/PhysRevLett.84.3760",
    journal = "Phys. Rev. Lett.",
    volume = "84",
    pages = "3760--3763",
    year = "2000"
}

@article{Kaplinghat:2015aga,
    author = "Kaplinghat, Manoj and Tulin, Sean and Yu, Hai-Bo",
    title = "{Dark Matter Halos as Particle Colliders: Unified Solution to Small-Scale Structure Puzzles from Dwarfs to Clusters}",
    eprint = "1508.03339",
    archivePrefix = "arXiv",
    primaryClass = "astro-ph.CO",
    doi = "10.1103/PhysRevLett.116.041302",
    journal = "Phys. Rev. Lett.",
    volume = "116",
    number = "4",
    pages = "041302",
    year = "2016"
}

@article{Newman:2012nw,
    author = "Newman, Andrew B. and Treu, Tommaso and Ellis, Richard S. and Sand, David J.",
    title = "{The Density Profiles of Massive, Relaxed Galaxy Clusters: II. Separating Luminous and Dark Matter in Cluster Cores}",
    eprint = "1209.1392",
    archivePrefix = "arXiv",
    primaryClass = "astro-ph.CO",
    doi = "10.1088/0004-637X/765/1/25",
    journal = "Astrophys. J.",
    volume = "765",
    pages = "25",
    year = "2013"
}

@article{KuziodeNaray:2007qi,
    author = "Kuzio de Naray, Rachel and McGaugh, Stacy S. and de Blok, W. J. G.",
    title = "{Mass Models for Low Surface Brightness Galaxies with High Resolution Optical Velocity Fields}",
    eprint = "0712.0860",
    archivePrefix = "arXiv",
    primaryClass = "astro-ph",
    doi = "10.1086/527543",
    journal = "Astrophys. J.",
    volume = "676",
    pages = "920--943",
    year = "2008"
}

@article{Oh:2010ea,
    author = "Oh, Se-Heon and de Blok, W. J. G. and Brinks, Elias and Walter, Fabian and Kennicutt, Jr, Robert C.",
    title = "{Dark and luminous matter in THINGS dwarf galaxies}",
    eprint = "1011.0899",
    archivePrefix = "arXiv",
    primaryClass = "astro-ph.CO",
    doi = "10.1088/0004-6256/141/6/193",
    journal = "Astron. J.",
    volume = "141",
    pages = "193",
    year = "2011"
}

@article{Garcia-Garcia:2024gzy,
    author = "Garc\'\i{}a-Garc\'\i{}a, Carlos and Zennaro, Matteo and Aric\`o, Giovanni and Alonso, David and Angulo, Raul E.",
    title = "{Cosmic shear with small scales: DES-Y3, KiDS-1000 and HSC-DR1}",
    eprint = "2403.13794",
    archivePrefix = "arXiv",
    primaryClass = "astro-ph.CO",
    doi = "10.1088/1475-7516/2024/08/024",
    journal = "JCAP",
    volume = "08",
    pages = "024",
    year = "2024"
}

@article{Scherer:2002tk,
    author = "Scherer, Stefan",
    editor = "Negele, John W. and Vogt, E. W.",
    title = "{Introduction to chiral perturbation theory}",
    eprint = "hep-ph/0210398",
    archivePrefix = "arXiv",
    reportNumber = "MKPH-T-02-09",
    journal = "Adv. Nucl. Phys.",
    volume = "27",
    pages = "277",
    year = "2003"
}

@article{Chu:2018fzy,
    author = "Chu, Xiaoyong and Garcia-Cely, Camilo and Murayama, Hitoshi",
    title = "{Velocity Dependence from Resonant Self-Interacting Dark Matter}",
    eprint = "1810.04709",
    archivePrefix = "arXiv",
    primaryClass = "hep-ph",
    reportNumber = "DESY-18-176",
    doi = "10.1103/PhysRevLett.122.071103",
    journal = "Phys. Rev. Lett.",
    volume = "122",
    number = "7",
    pages = "071103",
    year = "2019"
}

@article{Sokolenko:2018noz,
    author = "Sokolenko, Anastasia and Bondarenko, Kyrylo and Brinckmann, Thejs and Zavala, Jes\'us and Vogelsberger, Mark and Bringmann, Torsten and Boyarsky, Alexey",
    title = "{Towards an improved model of self-interacting dark matter haloes}",
    eprint = "1806.11539",
    archivePrefix = "arXiv",
    primaryClass = "astro-ph.CO",
    doi = "10.1088/1475-7516/2018/12/038",
    journal = "JCAP",
    volume = "12",
    pages = "038",
    year = "2018"
}

@article{Binder:2017rgn,
    author = "Binder, Tobias and Bringmann, Torsten and Gustafsson, Michael and Hryczuk, Andrzej",
    title = "{Early kinetic decoupling of dark matter: when the standard way of calculating the thermal relic density fails}",
    eprint = "1706.07433",
    archivePrefix = "arXiv",
    primaryClass = "astro-ph.CO",
    doi = "10.1103/PhysRevD.96.115010",
    journal = "Phys. Rev. D",
    volume = "96",
    number = "11",
    pages = "115010",
    year = "2017",
    note = "[Erratum: Phys.Rev.D 101, 099901 (2020)]"
}

@article{Binder:2021bmg,
    author = "Binder, Tobias and Bringmann, Torsten and Gustafsson, Michael and Hryczuk, Andrzej",
    title = "{Dark matter relic abundance beyond kinetic equilibrium}",
    eprint = "2103.01944",
    archivePrefix = "arXiv",
    primaryClass = "hep-ph",
    doi = "10.1140/epjc/s10052-021-09357-5",
    journal = "Eur. Phys. J. C",
    volume = "81",
    pages = "577",
    year = "2021"
}

@article{Saikawa:2020swg,
    author = "Saikawa, Ken'ichi and Shirai, Satoshi",
    title = "{Precise WIMP Dark Matter Abundance and Standard Model Thermodynamics}",
    eprint = "2005.03544",
    archivePrefix = "arXiv",
    primaryClass = "hep-ph",
    reportNumber = "IPMU20-0047, KANAZAWA-20-03, MPP-2020-62",
    doi = "10.1088/1475-7516/2020/08/011",
    journal = "JCAP",
    volume = "08",
    pages = "011",
    year = "2020"
}

@article{Li:2023puz,
    author = "Li, Shao-Ping and Xu, Xun-Jie",
    title = "{N$_{eff}$ constraints on light mediators coupled to neutrinos: the dilution-resistant effect}",
    eprint = "2307.13967",
    archivePrefix = "arXiv",
    primaryClass = "hep-ph",
    doi = "10.1007/JHEP10(2023)012",
    journal = "JHEP",
    volume = "10",
    pages = "012",
    year = "2023"
}

@article{Yanagida:1979as,
    author = "Yanagida, Tsutomu",
    editor = "Sawada, Osamu and Sugamoto, Akio",
    title = "{Horizontal gauge symmetry and masses of neutrinos}",
    reportNumber = "KEK-79-18-95",
    journal = "Conf. Proc. C",
    volume = "7902131",
    pages = "95--99",
    year = "1979"
}

@article{Gell-Mann:1979vob,
    author = "Gell-Mann, Murray and Ramond, Pierre and Slansky, Richard",
    title = "{Complex Spinors and Unified Theories}",
    eprint = "1306.4669",
    archivePrefix = "arXiv",
    primaryClass = "hep-th",
    reportNumber = "PRINT-80-0576",
    journal = "Conf. Proc. C",
    volume = "790927",
    pages = "315--321",
    year = "1979"
}

@article{Fukugita:1986hr,
    author = "Fukugita, M. and Yanagida, T.",
    title = "{Baryogenesis Without Grand Unification}",
    reportNumber = "RIFP-641",
    doi = "10.1016/0370-2693(86)91126-3",
    journal = "Phys. Lett. B",
    volume = "174",
    pages = "45--47",
    year = "1986"
}

@article{Buchmuller:2005eh,
    author = "Buchmuller, W. and Peccei, R. D. and Yanagida, T.",
    title = "{Leptogenesis as the origin of matter}",
    eprint = "hep-ph/0502169",
    archivePrefix = "arXiv",
    reportNumber = "DESY-05-031",
    doi = "10.1146/annurev.nucl.55.090704.151558",
    journal = "Ann. Rev. Nucl. Part. Sci.",
    volume = "55",
    pages = "311--355",
    year = "2005"
}

@article{Lin:2022xbu,
    author = "Lin, Weikang and Visinelli, Luca and Xu, Donglian and Yanagida, Tsutomu T.",
    title = "{Neutrino astronomy as a probe of physics beyond the Standard Model: Decay of sub-MeV B-L gauge boson dark matter}",
    eprint = "2202.04496",
    archivePrefix = "arXiv",
    primaryClass = "hep-ph",
    doi = "10.1103/PhysRevD.106.075011",
    journal = "Phys. Rev. D",
    volume = "106",
    number = "7",
    pages = "075011",
    year = "2022"
}

@article{Graham:2021ggy,
    author = "Graham, Matt and Hearty, Christopher and Williams, Mike",
    title = "{Searches for Dark Photons at Accelerators}",
    eprint = "2104.10280",
    archivePrefix = "arXiv",
    primaryClass = "hep-ph",
    doi = "10.1146/annurev-nucl-110320-051823",
    journal = "Ann. Rev. Nucl. Part. Sci.",
    volume = "71",
    pages = "37--58",
    year = "2021"
}

@article{Ferber:2023iso,
    author = "Ferber, Torben and Grohsjean, Alexander and Kahlhoefer, Felix",
    title = "{Dark Higgs bosons at colliders}",
    eprint = "2305.16169",
    archivePrefix = "arXiv",
    primaryClass = "hep-ph",
    reportNumber = "P3H-23-034, TTP23-018",
    doi = "10.1016/j.ppnp.2024.104105",
    journal = "Prog. Part. Nucl. Phys.",
    volume = "136",
    pages = "104105",
    year = "2024"
}

@article{Essig:2011nj,
    author = "Essig, Rouven and Mardon, Jeremy and Volansky, Tomer",
    title = "{Direct Detection of Sub-GeV Dark Matter}",
    eprint = "1108.5383",
    archivePrefix = "arXiv",
    primaryClass = "hep-ph",
    reportNumber = "SLAC-PUB-14538",
    doi = "10.1103/PhysRevD.85.076007",
    journal = "Phys. Rev. D",
    volume = "85",
    pages = "076007",
    year = "2012"
}

@article{Schumann:2019eaa,
    author = "Schumann, Marc",
    title = "{Direct Detection of WIMP Dark Matter: Concepts and Status}",
    eprint = "1903.03026",
    archivePrefix = "arXiv",
    primaryClass = "astro-ph.CO",
    doi = "10.1088/1361-6471/ab2ea5",
    journal = "J. Phys. G",
    volume = "46",
    number = "10",
    pages = "103003",
    year = "2019"
}

@article{Kamada:2016euw,
    author = "Kamada, Ayuki and Kaplinghat, Manoj and Pace, Andrew B. and Yu, Hai-Bo",
    title = "{How the Self-Interacting Dark Matter Model Explains the Diverse Galactic Rotation Curves}",
    eprint = "1611.02716",
    archivePrefix = "arXiv",
    primaryClass = "astro-ph.GA",
    doi = "10.1103/PhysRevLett.119.111102",
    journal = "Phys. Rev. Lett.",
    volume = "119",
    number = "11",
    pages = "111102",
    year = "2017"
}

@article{Gondolo:1990dk,
    author = "Gondolo, Paolo and Gelmini, Graciela",
    title = "{Cosmic abundances of stable particles: Improved analysis}",
    reportNumber = "UCLA-90-TEP-68",
    doi = "10.1016/0550-3213(91)90438-4",
    journal = "Nucl. Phys. B",
    volume = "360",
    pages = "145--179",
    year = "1991"
}

@article{Dolgov:2002wy,
    author = "Dolgov, A. D.",
    title = "{Neutrinos in cosmology}",
    eprint = "hep-ph/0202122",
    archivePrefix = "arXiv",
    reportNumber = "INFN-FE",
    doi = "10.1016/S0370-1573(02)00139-4",
    journal = "Phys. Rept.",
    volume = "370",
    pages = "333--535",
    year = "2002"
}

@article{Ibe:2018juk,
    author = "Ibe, Masahiro and Kamada, Ayuki and Kobayashi, Shin and Nakano, Wakutaka",
    title = "{Composite Asymmetric Dark Matter with a Dark Photon Portal}",
    eprint = "1805.06876",
    archivePrefix = "arXiv",
    primaryClass = "hep-ph",
    doi = "10.1007/JHEP11(2018)203",
    journal = "JHEP",
    volume = "11",
    pages = "203",
    year = "2018"
}

@article{Boehm:2013jpa,
    author = "Boehm, C{\'e}line and Dolan, Matthew J. and McCabe, Christopher",
    title = "{A Lower Bound on the Mass of Cold Thermal Dark Matter from Planck}",
    eprint = "1303.6270",
    archivePrefix = "arXiv",
    primaryClass = "hep-ph",
    reportNumber = "IPPP-13-18, DCPT-13-36",
    doi = "10.1088/1475-7516/2013/08/041",
    journal = "JCAP",
    volume = "08",
    pages = "041",
    year = "2013"
}

@article{Giovanetti:2021izc,
    author = "Giovanetti, Cara and Lisanti, Mariangela and Liu, Hongwan and Ruderman, Joshua T.",
    title = "{Joint Cosmic Microwave Background and Big Bang Nucleosynthesis Constraints on Light Dark Sectors with Dark Radiation}",
    eprint = "2109.03246",
    archivePrefix = "arXiv",
    primaryClass = "hep-ph",
    doi = "10.1103/PhysRevLett.129.021302",
    journal = "Phys. Rev. Lett.",
    volume = "129",
    number = "2",
    pages = "021302",
    year = "2022"
}

@article{Sabti:2021reh,
    author = "Sabti, Nashwan and Alvey, James and Escudero, Miguel and Fairbairn, Malcolm and Blas, Diego",
    title = "{Addendum: Refined bounds on MeV-scale thermal dark sectors from BBN and the CMB}",
    eprint = "2107.11232",
    archivePrefix = "arXiv",
    primaryClass = "hep-ph",
    doi = "10.1088/1475-7516/2021/08/A01",
    journal = "JCAP",
    volume = "08",
    pages = "A01",
    year = "2021"
}

@article{Chu:2022xuh,
    author = "Chu, Xiaoyong and Kuo, Jui-Lin and Pradler, Josef",
    title = "{Toward a full description of MeV dark matter decoupling: A self-consistent determination of relic abundance and Neff}",
    eprint = "2205.05714",
    archivePrefix = "arXiv",
    primaryClass = "hep-ph",
    doi = "10.1103/PhysRevD.106.055022",
    journal = "Phys. Rev. D",
    volume = "106",
    number = "5",
    pages = "055022",
    year = "2022"
}

@article{Planck:2018vyg,
    author = "Aghanim, N. and others",
    collaboration = "Planck",
    title = "{Planck 2018 results. VI. Cosmological parameters}",
    eprint = "1807.06209",
    archivePrefix = "arXiv",
    primaryClass = "astro-ph.CO",
    doi = "10.1051/0004-6361/201833910",
    journal = "Astron. Astrophys.",
    volume = "641",
    pages = "A6",
    year = "2020",
    note = "[Erratum: Astron.Astrophys. 652, C4 (2021)]"
}

@article{Verde:2019ivm,
    author = "Verde, L. and Treu, T. and Riess, A. G.",
    title = "{Tensions between the Early and the Late Universe}",
    eprint = "1907.10625",
    archivePrefix = "arXiv",
    primaryClass = "astro-ph.CO",
    doi = "10.1038/s41550-019-0902-0",
    journal = "Nature Astron.",
    volume = "3",
    pages = "891",
    year = "2019"
}

@article{deSalas:2016ztq,
    author = "de Salas, Pablo F. and Pastor, Sergio",
    title = "{Relic neutrino decoupling with flavour oscillations revisited}",
    eprint = "1606.06986",
    archivePrefix = "arXiv",
    primaryClass = "hep-ph",
    reportNumber = "IFIC-16-10, TTK-16-23",
    doi = "10.1088/1475-7516/2016/07/051",
    journal = "JCAP",
    volume = "07",
    pages = "051",
    year = "2016"
}

@article{Bennett:2020zkv,
    author = "Bennett, Jack J. and Buldgen, Gilles and De Salas, Pablo F. and Drewes, Marco and Gariazzo, Stefano and Pastor, Sergio and Wong, Yvonne Y. Y.",
    title = "{Towards a precision calculation of $N_{\rm eff}$ in the Standard Model II: Neutrino decoupling in the presence of flavour oscillations and finite-temperature QED}",
    eprint = "2012.02726",
    archivePrefix = "arXiv",
    primaryClass = "hep-ph",
    reportNumber = "CPPC-2020-10",
    doi = "10.1088/1475-7516/2021/04/073",
    journal = "JCAP",
    volume = "04",
    pages = "073",
    year = "2021"
}

@article{ParticleDataGroup:2024cfk,
    author = "Navas, S. and others",
    collaboration = "Particle Data Group",
    title = "{Review of particle physics}",
    doi = "10.1103/PhysRevD.110.030001",
    journal = "Phys. Rev. D",
    volume = "110",
    number = "3",
    pages = "030001",
    year = "2024"
}

@article{Matsumoto:2018acr,
    author = "Matsumoto, Shigeki and Tsai, Yue-Lin Sming and Tseng, Po-Yan",
    title = "{Light Fermionic WIMP Dark Matter with Light Scalar Mediator}",
    eprint = "1811.03292",
    archivePrefix = "arXiv",
    primaryClass = "hep-ph",
    reportNumber = "IPMU18-0180",
    doi = "10.1007/JHEP07(2019)050",
    journal = "JHEP",
    volume = "07",
    pages = "050",
    year = "2019"
}

@article{Kawasaki:1994af,
    author = "Kawasaki, M. and Moroi, T.",
    title = "{Gravitino production in the inflationary universe and the effects on big bang nucleosynthesis}",
    eprint = "hep-ph/9403364",
    archivePrefix = "arXiv",
    reportNumber = "ICRR-315-94-10, TU-457",
    doi = "10.1143/PTP.93.879",
    journal = "Prog. Theor. Phys.",
    volume = "93",
    pages = "879--900",
    year = "1995"
}

@article{Ellis:1990nb,
    author = "Ellis, John R. and Gelmini, G. B. and Lopez, Jorge L. and Nanopoulos, Dimitri V. and Sarkar, Subir",
    title = "{Astrophysical constraints on massive unstable neutral relic particles}",
    reportNumber = "CERN-TH-5853-90, CTP-TAMU-96-90, ACT-24",
    doi = "10.1016/0550-3213(92)90438-H",
    journal = "Nucl. Phys. B",
    volume = "373",
    pages = "399--437",
    year = "1992"
}

@article{Arguelles:2019ouk,
    author = {Arg{\"u}elles, Carlos A. and Diaz, Alejandro and Kheirandish, Ali and Olivares-Del-Campo, Andr{\'e}s and Safa, Ibrahim and Vincent, Aaron C.},
    title = "{Dark matter annihilation to neutrinos}",
    eprint = "1912.09486",
    archivePrefix = "arXiv",
    primaryClass = "hep-ph",
    doi = "10.1103/RevModPhys.93.035007",
    journal = "Rev. Mod. Phys.",
    volume = "93",
    number = "3",
    pages = "035007",
    year = "2021"
}

@article{Oman:2015xda,
    author = "Oman, Kyle A. and others",
    title = "{The unexpected diversity of dwarf galaxy rotation curves}",
    eprint = "1504.01437",
    archivePrefix = "arXiv",
    primaryClass = "astro-ph.GA",
    doi = "10.1093/mnras/stv1504",
    journal = "Mon. Not. Roy. Astron. Soc.",
    volume = "452",
    number = "4",
    pages = "3650--3665",
    year = "2015"
}

@article{Shalgar:2019rqe,
    author = "Shalgar, Shashank and Tamborra, Irene and Bustamante, Mauricio",
    title = "{Core-collapse supernovae stymie secret neutrino interactions}",
    eprint = "1912.09115",
    archivePrefix = "arXiv",
    primaryClass = "astro-ph.HE",
    doi = "10.1103/PhysRevD.103.123008",
    journal = "Phys. Rev. D",
    volume = "103",
    number = "12",
    pages = "123008",
    year = "2021"
}

@article{Benito:2020lgu,
    author = "Benito, Mar{\'\i}a and Iocco, Fabio and Cuoco, Alessandro",
    title = "{Uncertainties in the Galactic Dark Matter distribution: An update}",
    eprint = "2009.13523",
    archivePrefix = "arXiv",
    primaryClass = "astro-ph.GA",
    doi = "10.1016/j.dark.2021.100826",
    journal = "Phys. Dark Univ.",
    volume = "32",
    pages = "100826",
    year = "2021"
}

@article{Burkert:1995yz,
    author = "Burkert, A.",
    title = "{The Structure of dark matter halos in dwarf galaxies}",
    eprint = "astro-ph/9504041",
    archivePrefix = "arXiv",
    doi = "10.1086/309560",
    journal = "Astrophys. J. Lett.",
    volume = "447",
    pages = "L25",
    year = "1995"
}

@article{Benito:2019ngh,
    author = "Benito, Maria and Cuoco, Alessandro and Iocco, Fabio",
    title = "{Handling the Uncertainties in the Galactic Dark Matter Distribution for Particle Dark Matter Searches}",
    eprint = "1901.02460",
    archivePrefix = "arXiv",
    primaryClass = "astro-ph.GA",
    doi = "10.1088/1475-7516/2019/03/033",
    journal = "JCAP",
    volume = "03",
    pages = "033",
    year = "2019"
}

@article{Choi:2020kch,
    author = "Choi, Gongjun and Yanagida, Tsutomu T. and Yokozaki, Norimi",
    title = "{Feebly interacting $U (1)_{B-L}$ gauge boson warm dark matter and XENON1T anomaly}",
    eprint = "2007.04278",
    archivePrefix = "arXiv",
    primaryClass = "hep-ph",
    doi = "10.1016/j.physletb.2020.135836",
    journal = "Phys. Lett. B",
    volume = "810",
    pages = "135836",
    year = "2020"
}

@article{Follin:2015hya,
    author = "Follin, Brent and Knox, Lloyd and Millea, Marius and Pan, Zhen",
    title = "{First Detection of the Acoustic Oscillation Phase Shift Expected from the Cosmic Neutrino Background}",
    eprint = "1503.07863",
    archivePrefix = "arXiv",
    primaryClass = "astro-ph.CO",
    doi = "10.1103/PhysRevLett.115.091301",
    journal = "Phys. Rev. Lett.",
    volume = "115",
    number = "9",
    pages = "091301",
    year = "2015"
}

@article{Baumann:2017lmt,
    author = "Baumann, Daniel and Green, Daniel and Zaldarriaga, Matias",
    title = "{Phases of New Physics in the BAO Spectrum}",
    eprint = "1703.00894",
    archivePrefix = "arXiv",
    primaryClass = "astro-ph.CO",
    doi = "10.1088/1475-7516/2017/11/007",
    journal = "JCAP",
    volume = "11",
    pages = "007",
    year = "2017"
}

@article{Choi:2018gho,
    author = "Choi, Gongjun and Chiang, Chi-Ting and LoVerde, Marilena",
    title = "{Probing Decoupling in Dark Sectors with the Cosmic Microwave Background}",
    eprint = "1804.10180",
    archivePrefix = "arXiv",
    primaryClass = "astro-ph.CO",
    reportNumber = "YITP-SB-18-09",
    doi = "10.1088/1475-7516/2018/06/044",
    journal = "JCAP",
    volume = "06",
    pages = "044",
    year = "2018"
}

@article{Kreisch:2019yzn,
    author = "Kreisch, Christina D. and Cyr-Racine, Francis-Yan and Dor{\'e}, Olivier",
    title = "{Neutrino puzzle: Anomalies, interactions, and cosmological tensions}",
    eprint = "1902.00534",
    archivePrefix = "arXiv",
    primaryClass = "astro-ph.CO",
    doi = "10.1103/PhysRevD.101.123505",
    journal = "Phys. Rev. D",
    volume = "101",
    number = "12",
    pages = "123505",
    year = "2020"
}

@article{Ma:1995ey,
    author = "Ma, Chung-Pei and Bertschinger, Edmund",
    title = "{Cosmological perturbation theory in the synchronous and conformal Newtonian gauges}",
    eprint = "astro-ph/9506072",
    archivePrefix = "arXiv",
    doi = "10.1086/176550",
    journal = "Astrophys. J.",
    volume = "455",
    pages = "7--25",
    year = "1995"
}

@book{Dodelson:2003ft,
    author = "Dodelson, Scott",
    title = "{Modern Cosmology}",
    isbn = "978-0-12-219141-1",
    publisher = "Academic Press",
    address = "Amsterdam",
    year = "2003"
}

@article{Das:2020xke,
    author = "Das, Anirban and Ghosh, Subhajit",
    title = "{Flavor-specific interaction favors strong neutrino self-coupling in the early universe}",
    eprint = "2011.12315",
    archivePrefix = "arXiv",
    primaryClass = "astro-ph.CO",
    reportNumber = "SLAC-PUB-17547",
    doi = "10.1088/1475-7516/2021/07/038",
    journal = "JCAP",
    volume = "07",
    pages = "038",
    year = "2021"
}

@article{Blinov:2019gcj,
    author = "Blinov, Nikita and Kelly, Kevin James and Krnjaic, Gordan Z and McDermott, Samuel D",
    title = "{Constraining the Self-Interacting Neutrino Interpretation of the Hubble Tension}",
    eprint = "1905.02727",
    archivePrefix = "arXiv",
    primaryClass = "astro-ph.CO",
    reportNumber = "FERMILAB-PUB-19-175-A-T",
    doi = "10.1103/PhysRevLett.123.191102",
    journal = "Phys. Rev. Lett.",
    volume = "123",
    number = "19",
    pages = "191102",
    year = "2019"
}

@article{Grohs:2020xxd,
    author = "Grohs, E. and Fuller, George M. and Sen, Manibrata",
    title = "{Consequences of neutrino self interactions for weak decoupling and big bang nucleosynthesis}",
    eprint = "2002.08557",
    archivePrefix = "arXiv",
    primaryClass = "astro-ph.CO",
    doi = "10.1088/1475-7516/2020/07/001",
    journal = "JCAP",
    volume = "07",
    pages = "001",
    year = "2020"
}

@article{Huang:2021dba,
    author = "Huang, Guo-yuan and Rodejohann, Werner",
    title = "{Solving the Hubble tension without spoiling Big Bang Nucleosynthesis}",
    eprint = "2102.04280",
    archivePrefix = "arXiv",
    primaryClass = "hep-ph",
    doi = "10.1103/PhysRevD.103.123007",
    journal = "Phys. Rev. D",
    volume = "103",
    pages = "123007",
    year = "2021"
}

@article{Wilkinson:2014ksa,
    author = "Wilkinson, Ryan J. and Boehm, Celine and Lesgourgues, Julien",
    title = "{Constraining Dark Matter-Neutrino Interactions using the CMB and Large-Scale Structure}",
    eprint = "1401.7597",
    archivePrefix = "arXiv",
    primaryClass = "astro-ph.CO",
    reportNumber = "IPPP-14-03, DCPT-14-06, CERN-PH-TH-2014-013, LAPTH-006-14",
    doi = "10.1088/1475-7516/2014/05/011",
    journal = "JCAP",
    volume = "05",
    pages = "011",
    year = "2014"
}

@article{Escudero:2015yka,
    author = "Escudero, Miguel and Mena, Olga and Vincent, Aaron C. and Wilkinson, Ryan J. and B{\oe}hm, C{\'e}line",
    title = "{Exploring dark matter microphysics with galaxy surveys}",
    eprint = "1505.06735",
    archivePrefix = "arXiv",
    primaryClass = "astro-ph.CO",
    reportNumber = "IFIC-15-32, IPPP-15-29, DCPT-15-58",
    doi = "10.1088/1475-7516/2015/9/034",
    journal = "JCAP",
    volume = "09",
    pages = "034",
    year = "2015"
}

@article{Bertoni:2014mva,
    author = "Bertoni, Bridget and Ipek, Seyda and McKeen, David and Nelson, Ann E.",
    title = "{Constraints and consequences of reducing small scale structure via large dark matter-neutrino interactions}",
    eprint = "1412.3113",
    archivePrefix = "arXiv",
    primaryClass = "hep-ph",
    doi = "10.1007/JHEP04(2015)170",
    journal = "JHEP",
    volume = "04",
    pages = "170",
    year = "2015"
}

@article{Kelly:2018tyg,
    author = "Kelly, Kevin J. and Machado, Pedro A. N.",
    title = "{Multimessenger Astronomy and New Neutrino Physics}",
    eprint = "1808.02889",
    archivePrefix = "arXiv",
    primaryClass = "hep-ph",
    reportNumber = "FERMILAB-PUB-18-374-T, NUHEP-TH/18-07",
    doi = "10.1088/1475-7516/2018/10/048",
    journal = "JCAP",
    volume = "10",
    pages = "048",
    year = "2018"
}

@article{Bustamante:2020mep,
    author = "Bustamante, Mauricio and Rosenstr{\o}m, Charlotte and Shalgar, Shashank and Tamborra, Irene",
    title = "{Bounds on secret neutrino interactions from high-energy astrophysical neutrinos}",
    eprint = "2001.04994",
    archivePrefix = "arXiv",
    primaryClass = "astro-ph.HE",
    doi = "10.1103/PhysRevD.101.123024",
    journal = "Phys. Rev. D",
    volume = "101",
    number = "12",
    pages = "123024",
    year = "2020"
}

@article{Petropavlova:2024wia,
    author = "Petropavlova, Maria",
    title = "{Constraints on Neutrino Self Interactions from Multi-messenger neutrinos scattering on C$\nu$B}",
    eprint = "2505.14332",
    archivePrefix = "arXiv",
    primaryClass = "hep-ph",
    doi = "10.22323/1.476.0185",
    journal = "PoS",
    volume = "ICHEP2024",
    pages = "185",
    year = "2025"
}

@article{Cline:2023tkp,
    author = "Cline, James M. and Puel, Matteo",
    title = "{NGC 1068 constraints on neutrino-dark matter scattering}",
    eprint = "2301.08756",
    archivePrefix = "arXiv",
    primaryClass = "hep-ph",
    doi = "10.1088/1475-7516/2023/06/004",
    journal = "JCAP",
    volume = "06",
    pages = "004",
    year = "2023"
}

@article{Fujiwara:2023lsv,
    author = "Fujiwara, Motoko and Herrera, Gonzalo",
    title = "{Tidal disruption events and dark matter scatterings with neutrinos and photons}",
    eprint = "2312.11670",
    archivePrefix = "arXiv",
    primaryClass = "hep-ph",
    doi = "10.1016/j.physletb.2024.138573",
    journal = "Phys. Lett. B",
    volume = "851",
    pages = "138573",
    year = "2024"
}

@article{Heston:2024ljf,
    author = "Heston, Sean and Horiuchi, Shunsaku and Shirai, Satoshi",
    title = "{Constraining neutrino-DM interactions with Milky~Way dwarf spheroidals and supernova neutrinos}",
    eprint = "2402.08718",
    archivePrefix = "arXiv",
    primaryClass = "hep-ph",
    doi = "10.1103/PhysRevD.110.023004",
    journal = "Phys. Rev. D",
    volume = "110",
    number = "2",
    pages = "023004",
    year = "2024"
}

@article{XENON:2019zpr,
    author = "Aprile, E. and others",
    collaboration = "XENON",
    title = "{Search for Light Dark Matter Interactions Enhanced by the Migdal Effect or Bremsstrahlung in XENON1T}",
    eprint = "1907.12771",
    archivePrefix = "arXiv",
    primaryClass = "hep-ex",
    doi = "10.1103/PhysRevLett.123.241803",
    journal = "Phys. Rev. Lett.",
    volume = "123",
    number = "24",
    pages = "241803",
    year = "2019"
}

@article{COHERENT:2017ipa,
    author = "Akimov, D. and others",
    collaboration = "COHERENT",
    title = "{Observation of Coherent Elastic Neutrino-Nucleus Scattering}",
    eprint = "1708.01294",
    archivePrefix = "arXiv",
    primaryClass = "nucl-ex",
    doi = "10.1126/science.aao0990",
    journal = "Science",
    volume = "357",
    number = "6356",
    pages = "1123--1126",
    year = "2017"
}

@article{COHERENT:2020iec,
    author = "Akimov, D. and others",
    collaboration = "COHERENT",
    title = "{First Measurement of Coherent Elastic Neutrino-Nucleus Scattering on Argon}",
    eprint = "2003.10630",
    archivePrefix = "arXiv",
    primaryClass = "nucl-ex",
    doi = "10.1103/PhysRevLett.126.012002",
    journal = "Phys. Rev. Lett.",
    volume = "126",
    number = "1",
    pages = "012002",
    year = "2021"
}

@article{COHERENT:2024axu,
    author = "Adamski, S. and others",
    collaboration = "COHERENT",
    title = "{First detection of coherent elastic neutrino-nucleus scattering on germanium}",
    eprint = "2406.13806",
    archivePrefix = "arXiv",
    primaryClass = "hep-ex",
    month = "6",
    year = "2024"
}

@article{Ackermann:2025obx,
    author = "Ackermann, N. and others",
    title = "{Direct observation of coherent elastic antineutrino{\textendash}nucleus scattering}",
    eprint = "2501.05206",
    archivePrefix = "arXiv",
    primaryClass = "hep-ex",
    doi = "10.1038/s41586-025-09322-2",
    journal = "Nature",
    volume = "643",
    number = "8074",
    pages = "1229--1233",
    year = "2025"
}

@article{Chattaraj:2025fvx,
    author = "Chattaraj, Ayan and Majumdar, Anirban and Srivastava, Rahul",
    title = "{Probing standard model and beyond with reactor CE{\ensuremath{\nu}}NS data of CONUS+ experiment}",
    eprint = "2501.12441",
    archivePrefix = "arXiv",
    primaryClass = "hep-ph",
    doi = "10.1016/j.physletb.2025.139438",
    journal = "Phys. Lett. B",
    volume = "864",
    pages = "139438",
    year = "2025"
}

@article{AtzoriCorona:2025ygn,
    author = "Atzori Corona, M. and Cadeddu, M. and Cargioli, N. and Dordei, F. and Giunti, C.",
    title = "{Reactor antineutrinos CE{\ensuremath{\nu}}NS on germanium: CONUS+ and TEXONO as a new gateway to SM and BSM physics}",
    eprint = "2501.18550",
    archivePrefix = "arXiv",
    primaryClass = "hep-ph",
    doi = "10.1103/n563-8v8d",
    journal = "Phys. Rev. D",
    volume = "112",
    number = "1",
    pages = "015007",
    year = "2025"
}

@article{TEXONO:2025sub,
    author = "Karada{\u{g}}, S. and others",
    collaboration = "TEXONO",
    title = "{Constraints on new physics with light mediators and generalized neutrino interactions via coherent elastic neutrino nucleus scattering}",
    eprint = "2502.20007",
    archivePrefix = "arXiv",
    primaryClass = "hep-ex",
    doi = "10.1103/63xf-t6fz",
    journal = "Phys. Rev. D",
    volume = "112",
    number = "3",
    pages = "035038",
    year = "2025"
}

@article{Beacham:2019nyx,
    author = "Beacham, J. and others",
    title = "{Physics Beyond Colliders at CERN: Beyond the Standard Model Working Group Report}",
    eprint = "1901.09966",
    archivePrefix = "arXiv",
    primaryClass = "hep-ex",
    reportNumber = "CERN-PBC-REPORT-2018-007",
    doi = "10.1088/1361-6471/ab4cd2",
    journal = "J. Phys. G",
    volume = "47",
    number = "1",
    pages = "010501",
    year = "2020"
}

@article{Goudzovski:2022vbt,
    author = "Goudzovski, Evgueni and others",
    title = "{New physics searches at kaon and hyperon factories}",
    eprint = "2201.07805",
    archivePrefix = "arXiv",
    primaryClass = "hep-ph",
    reportNumber = "FERMILAB-PUB-22-057-T",
    doi = "10.1088/1361-6633/ac9cee",
    journal = "Rept. Prog. Phys.",
    volume = "86",
    number = "1",
    pages = "016201",
    year = "2023"
}

@article{NA62:2025upx,
    author = "Cortina Gil, Eduardo and others",
    collaboration = "NA62",
    title = "{Searches for hidden sectors using $K^+\to\pi^+X$ decays}",
    eprint = "2507.17286",
    archivePrefix = "arXiv",
    primaryClass = "hep-ex",
    reportNumber = "CERN-EP-2025-167",
    month = "7",
    year = "2025"
}

@article{NA62:2019meo,
    author = "Cortina Gil, Eduardo and others",
    collaboration = "NA62",
    title = "{Search for production of an invisible dark photon in $\pi^0$ decays}",
    eprint = "1903.08767",
    archivePrefix = "arXiv",
    primaryClass = "hep-ex",
    reportNumber = "CERN-EP-2019-048",
    doi = "10.1007/JHEP05(2019)182",
    journal = "JHEP",
    volume = "05",
    pages = "182",
    year = "2019"
}

@article{NA62:2020pwi,
    author = "Cortina Gil, Eduardo and others",
    collaboration = "NA62",
    title = "{Search for $\pi^0$ decays to invisible particles}",
    eprint = "2010.07644",
    archivePrefix = "arXiv",
    primaryClass = "hep-ex",
    reportNumber = "CERN-EP-2020-193",
    doi = "10.1007/JHEP02(2021)201",
    journal = "JHEP",
    volume = "02",
    pages = "201",
    year = "2021"
}

@article{Preskill:1982,
    author = "Preskill, Jhon and Wise, Mark B. and Wilczek, Frank",
    title = "{Cosmology of the Invisible Axion}",
    reportNumber = "HUTP-82-A048",
    doi = "10.1016/0370-2693(83)90637-8",
    journal = "Phys. Lett. B",
    volume = "120",
    pages = "127--132",
    year = "1983"
}

@article{Dodelson:1994,
   title={Sterile neutrinos as dark matter},
   eprint = "hep-ph/9303287",
   archivePrefix = "arXiv",
   primaryClass = "hep-ph",
   volume={72},
   ISSN={0031-9007},
   url={http://dx.doi.org/10.1103/PhysRevLett.72.17},
   DOI={10.1103/physrevlett.72.17},
   number={1},
   journal={Phys. Rev. Lett.},
   publisher={American Physical Society (APS)},
   author={Dodelson, Scott and Widrow, Lawrence M.},
   year={1994},
   month=jan, pages={17–20} 
}

@article{Chapline:1975,
    author = "Chapline, George F.",
    title = "{Cosmological effects of primodial black holes}",
    doi = "10.1038/253251a0",
    journal = "Nature",
    volume = "253",
    pages = "251--252",
    year = "1975"
}

@article{Bertone:2005,
   title="{Particle dark matter: evidence, candidates and constraints}",
   volume={405},
   ISSN={0370-1573},
   url={http://dx.doi.org/10.1016/j.physrep.2004.08.031},
   DOI={10.1016/j.physrep.2004.08.031},
   number={5–6},
   journal={Physics Reports},
   publisher={Elsevier BV},
   author={Bertone, Gianfranco and Hooper, Dan and Silk, Joseph},
   year={2005},
   pages={279–390},
   eprint = "arXiv:hep-ph/0404175",
   archivePrefix = "arXiv",
   primaryClass = "hep-ph"
}

@article{Feng:2010,
   title="{Dark Matter Candidates from Particle Physics and Methods of Detection}",
   volume={48},
   ISSN={1545-4282},
   url={http://dx.doi.org/10.1146/annurev-astro-082708-101659},
   DOI={10.1146/annurev-astro-082708-101659},
   number={1},
   journal={Annual Review of Astronomy and Astrophysics},
   publisher={Annual Reviews},
   author={Feng, Jonathan L.},
   year={2010},
   pages={495–545},
   eprint = "1003.0904",
   archivePrefix = "arXiv",
   primaryClass = "astro-ph"
}

@article{Redondo:2009,
   title={Massive hidden photons as lukewarm dark matter},
   volume={2009},
   ISSN={1475-7516},
   url={http://dx.doi.org/10.1088/1475-7516/2009/02/005},
   DOI={10.1088/1475-7516/2009/02/005},
   number={02},
   journal={JCAP},
   publisher={IOP Publishing},
   author={Redondo, Javier and Postma, Marieke},
   year={2009},
   pages={005–005},
   eprint = "0811.0326",
   archivePrefix = "arXiv",
   primaryClass = "hep-ph"
}

@article{Lee:1977,
  author  = {Lee, Benjamin W. and Weinberg, Steven},
  title   = "{Cosmological Lower Bound on Heavy-Neutrino Masses}",
  journal = {Phys. Rev. Lett.},
  volume  = {39},
  pages   = {165--168},
  year    = {1977},
  doi     = {10.1103/PhysRevLett.39.165}
}

@article{Hochberg:2014,
   title="{Mechanism for Thermal Relic Dark Matter of Strongly Interacting Massive Particles}",
   eprint = "1402.5143",
   archivePrefix = "arXiv",
   primaryClass = "hep-ph",
   volume={113},
   ISSN={1079-7114},
   url={http://dx.doi.org/10.1103/PhysRevLett.113.171301},
   DOI={10.1103/physrevlett.113.171301},
   number={17},
   journal={Phys. Rev. Lett.},
   publisher={American Physical Society (APS)},
   author={Hochberg, Yonit and Kuflik, Eric and Volansky, Tomer and Wacker, Jay G.},
   year={2014},
}

@article{Goldberg:1983,
  author        = {Goldberg, H.},
  title         = "{Constraint on the Photino Mass from Cosmology}",
  journal       = {Phys. Rev. Lett.},
  volume        = {50},
  pages         = {1419--1422},
  year          = {1983},
  doi           = {10.1103/PhysRevLett.50.1419},
  reportnumber  = {NUB-2592}
}

@article{Ellis:1984,
  author        = {Ellis, John R. and Hagelin, J. S. and Nanopoulos, Dimitri V. and Olive, Keith A. and Srednicki, M.},
  title         = "{Supersymmetric Relics from the Big Bang}",
  journal       = {Nucl. Phys. B},
  volume        = {238},
  pages         = {453--476},
  year          = {1984},
  doi           = {10.1016/0550-3213(84)90461-9},
  reportnumber  = {SLAC-PUB-3171}
}

@article{Ma:2006,
   title="{Verifiable radiative seesaw mechanism of neutrino mass and dark matter}",
   eprint = "hep-ph/0601225",
   archivePrefix = "arXiv",
   primaryClass = "hep-ph",
   volume={73},
   ISSN={1550-2368},
   url={http://dx.doi.org/10.1103/PhysRevD.73.077301},
   DOI={10.1103/physrevd.73.077301},
   number={7},
   journal={Phys. Rev. D},
   publisher={American Physical Society (APS)},
   author={Ma, Ernest},
   year={2006},
}

@article{Griest:1990,
  author        = {Griest, Kim and Kamionkowski, Marc},
  title         = "{Unitarity Limits on the Mass and Radius of Dark Matter Particles}",
  journal       = {Phys. Rev. Lett.},
  volume        = {64},
  pages         = {615--618},
  year          = {1990},
  doi           = {10.1103/PhysRevLett.64.615},
  reportnumber  = {CFPA-TH-89-013}
}

@article{Slatyer:2016,
   title="{Indirect dark matter signatures in the cosmic dark ages. {I}. Generalizing the bound on s-wave dark matter annihilation from Planck results}",
   volume={93},
   ISSN={2470-0029},
   url={http://dx.doi.org/10.1103/PhysRevD.93.023527},
   DOI={10.1103/physrevd.93.023527},
   number={2},
   journal={Phys. Rev. D},
   publisher={American Physical Society (APS)},
   author={Slatyer, Tracy R.},
   year={2016},
}

@article{Kawasaki:2021,
   title="{Revisiting CMB constraints on dark matter annihilation}",
   volume={2021},
   ISSN={1475-7516},
   url={http://dx.doi.org/10.1088/1475-7516/2021/12/015},
   DOI={10.1088/1475-7516/2021/12/015},
   number={12},
   journal={Journal of Cosmology and Astroparticle Physics},
   publisher={IOP Publishing},
   author={Kawasaki, Masahiro and Nakatsuka, Hiromasa and Nakayama, Kazunori and Sekiguchi, Toyokazu},
   year={2021},
   pages={015} 
}

@article{Drees:2022,
   title="{$U(1)_{L_\mu-L_\tau}$ for Light Dark Matter, $g_\mu-2$, the $511$ keV excess and the Hubble Tension}",
   volume={827},
   ISSN={0370-2693},
   url={http://dx.doi.org/10.1016/j.physletb.2022.136948},
   DOI={10.1016/j.physletb.2022.136948},
   journal={Physics Letters B},
   publisher={Elsevier BV},
   author={Drees, Manuel and Zhao, Wenbin},
   year={2022},
   pages={136948} }

@article{Herms:2023,
   title="{Light neutrinophilic dark matter from a scotogenic model}",
   eprint = "2307.15760",
   archivePrefix = "arXiv",
   primaryClass = "hep-ph",
   volume={845},
   ISSN={0370-2693},
   url={http://dx.doi.org/10.1016/j.physletb.2023.138167},
   DOI={10.1016/j.physletb.2023.138167},
   journal={Phys. Rev. B},
   publisher={Elsevier BV},
   author={Herms, Johannes and Jana, Sudip and P.K., Vishnu and Saad, Shaikh},
   year={2023},
   pages={138167} 
}

@article{B_hm:2004,
   title="{Scalar dark matter candidates}",
   eprint = "hep-ph/0305261v2",
   archivePrefix = "arXiv",
   primaryClass = "hep-ph",
   volume={683},
   ISSN={0550-3213},
   url={http://dx.doi.org/10.1016/j.nuclphysb.2004.01.015},
   DOI={10.1016/j.nuclphysb.2004.01.015},
   number={1–2},
   journal={Nucl. Phys. B},
   publisher={Elsevier BV},
   author={Bœhm, C. and Fayet, P.},
   year={2004},
   pages={219–263} 
}

@article{Bernreuther_2021,
   title="{Resonant sub-GeV Dirac dark matter}",
   eprint = "2010.14522",
   archivePrefix = "arXiv",
   primaryClass = "hep-ph",
   volume={2021},
   ISSN={1475-7516},
   url={http://dx.doi.org/10.1088/1475-7516/2021/03/040},
   DOI={10.1088/1475-7516/2021/03/040},
   number={03},
   journal={Journal of Cosmology and Astroparticle Physics},
   publisher={IOP Publishing},
   author={Bernreuther, Elias and Heeba, Saniya and Kahlhoefer, Felix},
   year={2021},
   month=mar, pages={040} }

@article{Griest:1991,
  author        = {Griest, Kim and Seckel, David},
  title         = "{Three exceptions in the calculation of relic abundances}",
  journal       = {Phys. Rev. D},
  volume        = {43},
  pages         = {3191--3203},
  year          = {1991},
  doi           = {10.1103/PhysRevD.43.3191},
  reportnumber  = {CFPA-TH-90-001A}
}

@article{Kuflik:2016,
   title="{Elastically Decoupling Dark Matter}",
   volume={116},
   ISSN={1079-7114},
   url={http://dx.doi.org/10.1103/PhysRevLett.116.221302},
   DOI={10.1103/physrevlett.116.221302},
   number={22},
   eprint = "1512.04545 ",
   archivePrefix = "arXiv",
   primaryClass = "hep-ph",
   journal={Phys. Rev. Lett.},
   publisher={American Physical Society (APS)},
   author={Kuflik, Eric and Perelstein, Maxim and Lorier, Nicolas Rey-Le and Tsai, Yu-Dai},
   year={2016},
   }

@article{Borexino:2019wln,
    author = "Agostini, M. and others",
    collaboration = "Borexino",
    title = "{Search for low-energy neutrinos from astrophysical sources with Borexino}",
    eprint = "1909.02422",
    archivePrefix = "arXiv",
    primaryClass = "hep-ex",
    reportNumber = "FERMILAB-PUB-21-152-AE",
    doi = "10.1016/j.astropartphys.2020.102509",
    journal = "Astropart. Phys.",
    volume = "125",
    pages = "102509",
    year = "2021"
}

@article{Watanabe:2025pvc,
    author = "Watanabe, Yu and Matsumoto, Shigeki and Karwin, Christopher M. and Melia, Tom and Negro, Michela and Siegert, Thomas and Watanabe, Yuki and Yoneda, Hiroki and Takahashi, Tadayuki",
    title = "{Sub-GeV dark matter and MeV gamma-ray detection with COSI}",
    eprint = "2504.11810",
    archivePrefix = "arXiv",
    primaryClass = "hep-ph",
    doi = "10.1007/JHEP09(2025)078",
    journal = "JHEP",
    volume = "09",
    pages = "078",
    year = "2025"
}

@article{KamLAND:2011bnd,
    author = "Gando, A. and others",
    collaboration = "KamLAND",
    title = "{A study of extraterrestrial antineutrino sources with the KamLAND detector}",
    eprint = "1105.3516",
    archivePrefix = "arXiv",
    primaryClass = "astro-ph.HE",
    doi = "10.1088/0004-637X/745/2/193",
    journal = "Astrophys. J.",
    volume = "745",
    pages = "193",
    year = "2012"
}

@phdthesis{Wan:2018ndu,
    author = "Wan, Linyan",
    title = "{Experimental Studies on Low Energy Electron Antineutrinos and Related Physics}",
    school = "Tsinghua U., Beijing",
    year = "2018"
}

@article{Rosenbluth:1957zz,
    author = "Rosenbluth, Marshall N. and MacDonald, William M. and Judd, David L.",
    title = "{Fokker-Planck Equation for an Inverse-Square Force}",
    doi = "10.1103/PhysRev.107.1",
    journal = "Phys. Rev.",
    volume = "107",
    pages = "1--6",
    year = "1957"
}

@article{Bringmann:2006mu,
    author = "Bringmann, Torsten and Hofmann, Stefan",
    title = "{Thermal decoupling of WIMPs from first principles}",
    eprint = "hep-ph/0612238",
    archivePrefix = "arXiv",
    doi = "10.1088/1475-7516/2007/04/016",
    journal = "JCAP",
    volume = "04",
    pages = "016",
    year = "2007",
    note = "[Erratum: JCAP 03, E02 (2016)]"
}

@article{Fayet:2016nyc,
    author = "Fayet, Pierre",
    title = "{The light $U$ boson as the mediator of a new force, coupled to a combination of $Q,B,L$ and dark matter}",
    eprint = "1611.05357",
    archivePrefix = "arXiv",
    primaryClass = "hep-ph",
    reportNumber = "LPTENS-16-07",
    doi = "10.1140/epjc/s10052-016-4568-9",
    journal = "Eur. Phys. J. C",
    volume = "77",
    number = "1",
    pages = "53",
    year = "2017"
}

@article{Kamionkowski:1990,
  author       = {Kamionkowski, Marc and Turner, Michael S.},
  title        = "{Thermal relics: Do we know their abundances?}",
  journal      = {Phys. Rev. D},
  volume       = {42},
  number       = {10},
  pages        = {3310--3320},
  year         = {1990},
  doi          = {10.1103/PhysRevD.42.3310}
}

@article{Kane:2015,
   title="{Cosmological moduli and the post-inflationary universe: A critical review}",
   volume={24},
   ISSN={1793-6594},
   url={http://dx.doi.org/10.1142/S0218271815300220},
   DOI={10.1142/s0218271815300220},
   number={08},
   journal={Int. J. Mod. Phys. D},
   publisher={World Scientific Pub Co Pte Lt},
   author={Kane, Gordon and Sinha, Kuver and Watson, Scott},
   year={2015},
  pages={1530022} }

@article{Jungman:1996,
   title={Supersymmetric dark matter},
   volume={267},
   ISSN={0370-1573},
   url={http://dx.doi.org/10.1016/0370-1573(95)00058-5},
   DOI={10.1016/0370-1573(95)00058-5},
   number={5–6},
   journal={Physics Reports},
   publisher={Elsevier BV},
   author={Jungman, Gerard and Kamionkowski, Marc and Griest, Kim},
   year={1996},
   pages={195–373},
   eprint  = {hep-ph/9506380},
   archivePrefix={arXiv},
   primaryClass={hep-ph},
}

@misc{essig2013darksectorsnewlight,
      title="{Dark Sectors and New, Light, Weakly-Coupled Particles}", 
      author={R. Essig et al.},
      year={2013},
      eprint={1311.0029},
      archivePrefix={arXiv},
      primaryClass={hep-ph},
      url={https://arxiv.org/abs/1311.0029}, 
}

@misc{alexander2016darksectors2016workshop,
      title="{Dark Sectors 2016 Workshop: Community Report}", 
      author={Jim Alexander et al.},
      year={2016},
      eprint={1608.08632},
      archivePrefix={arXiv},
      primaryClass={hep-ph},
      url={https://arxiv.org/abs/1608.08632}, 
}

@article{McKeown:2021sob,
    author = "McKeown, Daniel and Bullock, James S. and Mercado, Francisco J. and Hafen, Zachary and Boylan-Kolchin, Michael and Wetzel, Andrew and Necib, Lina and Hopkins, Philip F. and Yu, Sijie",
    title = "{Amplified J-factors in the Galactic Centre for velocity-dependent dark matter annihilation in FIRE simulations}",
    eprint = "2111.03076",
    archivePrefix = "arXiv",
    primaryClass = "astro-ph.GA",
    doi = "10.1093/mnras/stac966",
    journal = "Mon. Not. Roy. Astron. Soc.",
    volume = "513",
    number = "1",
    pages = "55--70",
    year = "2022"
}

@article{Lacroix:2020lhn,
    author = "Lacroix, Thomas and N{\'u}{\~n}ez-Casti{\~n}eyra, Arturo and Stref, Martin and Lavalle, Julien and Nezri, Emmanuel",
    title = "{Predicting the dark matter velocity distribution in galactic structures: tests against hydrodynamic cosmological simulations}",
    eprint = "2005.03955",
    archivePrefix = "arXiv",
    primaryClass = "astro-ph.GA",
    reportNumber = "LUPM:20-024, IFT-UAM/CSIC-20-64",
    doi = "10.1088/1475-7516/2020/10/031",
    journal = "JCAP",
    volume = "10",
    pages = "031",
    year = "2020"
}

@misc{esseili2025hidinglightvectorboson,
      title="{Hiding a Light Vector Boson from Terrestrial Experiments: A Chargephobic Dark Photon}", 
      author={Haidar Esseili and Graham D. Kribs},
      year={2025},
      eprint={2512.10916},
      archivePrefix={arXiv},
      primaryClass={hep-ph},
      url={https://arxiv.org/abs/2512.10916}, 
}

@article{Bell_2025,
   title="{Neutrino portals to MeV WIMPs with $s$ -channel mediators}",
   volume={111},
   ISSN={2470-0029},
   url={http://dx.doi.org/10.1103/PhysRevD.111.055020},
   DOI={10.1103/physrevd.111.055020},
   number={5},
   journal={Phys. Rev. D},
   publisher={American Physical Society (APS)},
   author={Bell, Nicole F. and Dolan, Matthew J. and Ghosh, Avirup and Virgato, Michael},
   year={2025},
   eprint  = {2412.02994},
   archivePrefix={arXiv},
   primaryClass={hep-ph},
}

@article{Abdallah:2021NeutrinophilicZp,
  author       = {Abdallah, Waleed and Barik, Anjan Kumar and Rai, Santosh Kumar and Samui, Tousik},
  title        = "{Search for a light {$Z^\prime$} at the LHC in a neutrinophilic {$U(1)$} model}",
  journal      = {Phys. Rev. D},
  publisher={American Physical Society (APS)},
  SSN={2470-0029},
  volume       = {104},
  number       = {9},
  year         = {2021},
  doi          = {10.1103/PhysRevD.104.095031}
}

\end{document}